\documentclass[twocolumn]{aastex631}
\usepackage{multirow}
\usepackage{amsmath}

\graphicspath{{./}{figures/}}

\begin{document}

\title{Measurements of the Temperature and E-mode Polarization of the Cosmic Microwave Background from the Full 500-square-degree SPTpol Dataset}

\shortauthors{T.-L.~Chou, et al.}
\author[0000-0002-3091-8790]{T.-L.~Chou} \affiliation{Department of Astronomy and Astrophysics, University of Chicago, 5640 South Ellis Avenue, Chicago, IL, USA 60637} \affiliation{Kavli Institute for Cosmological Physics, University of Chicago, 5640 South Ellis Avenue, Chicago, IL, USA 60637}
\author{P.~A.~R.~Ade} \affiliation{Cardiff University, Cardiff CF10 3XQ, United Kingdom}
\author[0000-0002-4435-4623]{A.~J.~Anderson} \affiliation{Fermi National Accelerator Laboratory, MS209, P.O. Box 500, Batavia, IL 60510}
\author{J.~E.~Austermann} \affiliation{NIST Quantum Devices Group, 325 Broadway Mailcode 817.03, Boulder, CO, USA 80305} \affiliation{Department of Physics, University of Colorado, Boulder, CO, USA 80309}
\author[0000-0001-6899-1873]{L.~Balkenhol} \affiliation{Sorbonne Universit'e, CNRS, UMR 7095, Institut d'Astrophysique de Paris, 98 bis bd Arago, 75014 Paris, France}
\author{J.~A.~Beall} \affiliation{NIST Quantum Devices Group, 325 Broadway Mailcode 817.03, Boulder, CO, USA 80305}
\author[0000-0001-5868-0748]{A.~N.~Bender} \affiliation{High Energy Physics Division, Argonne National Laboratory, 9700 S. Cass Avenue, Argonne, IL, USA 60439} \affiliation{Kavli Institute for Cosmological Physics, University of Chicago, 5640 South Ellis Avenue, Chicago, IL, USA 60637}
\author[0000-0002-5108-6823]{B.~A.~Benson} \affiliation{Fermi National Accelerator Laboratory, MS209, P.O. Box 500, Batavia, IL 60510} \affiliation{Kavli Institute for Cosmological Physics, University of Chicago, 5640 South Ellis Avenue, Chicago, IL, USA 60637} \affiliation{Department of Astronomy and Astrophysics, University of Chicago, 5640 South Ellis Avenue, Chicago, IL, USA 60637}
\author[0000-0003-4847-3483]{F.~Bianchini} \affiliation{SLAC National Accelerator Laboratory, 2575 Sand Hill Road, Menlo Park, CA 94025}
\author[0000-0001-7665-5079]{L.~E.~Bleem} \affiliation{High Energy Physics Division, Argonne National Laboratory, 9700 S. Cass Avenue, Argonne, IL, USA 60439} \affiliation{Kavli Institute for Cosmological Physics, University of Chicago, 5640 South Ellis Avenue, Chicago, IL, USA 60637}
\author{J.~E.~Carlstrom} \affiliation{Kavli Institute for Cosmological Physics, University of Chicago, 5640 South Ellis Avenue, Chicago, IL, USA 60637} \affiliation{Department of Physics, University of Chicago, 5640 South Ellis Avenue, Chicago, IL, USA 60637} \affiliation{High Energy Physics Division, Argonne National Laboratory, 9700 S. Cass Avenue, Argonne, IL, USA 60439} \affiliation{Department of Astronomy and Astrophysics, University of Chicago, 5640 South Ellis Avenue, Chicago, IL, USA 60637} \affiliation{Enrico Fermi Institute, University of Chicago, 5640 South Ellis Avenue, Chicago, IL, USA 60637}
\author{C.~L.~Chang} \affiliation{Kavli Institute for Cosmological Physics, University of Chicago, 5640 South Ellis Avenue, Chicago, IL, USA 60637} \affiliation{High Energy Physics Division, Argonne National Laboratory, 9700 S. Cass Avenue, Argonne, IL, USA 60439} \affiliation{Department of Astronomy and Astrophysics, University of Chicago, 5640 South Ellis Avenue, Chicago, IL, USA 60637}
\author{P.~Chaubal} \affiliation{School of Physics, University of Melbourne, Parkville, VIC 3010, Australia}
\author{H.~C.~Chiang} \affiliation{Department of Physics, McGill University, 3600 Rue University, Montreal, Quebec H3A 2T8, Canada} \affiliation{School of Mathematics, Statistics \& Computer Science, University of KwaZulu-Natal, Durban, South Africa}
\author{R.~Citron} \affiliation{University of Chicago, 5640 South Ellis Avenue, Chicago, IL, USA 60637}
\author{C.~Corbett~Moran} \affiliation{Jet Propulsion Laboratory, Pasadena, CA 91109, USA}
\author[0000-0001-9000-5013]{T.~M.~Crawford} \affiliation{Kavli Institute for Cosmological Physics, University of Chicago, 5640 South Ellis Avenue, Chicago, IL, USA 60637} \affiliation{Department of Astronomy and Astrophysics, University of Chicago, 5640 South Ellis Avenue, Chicago, IL, USA 60637}
\author{A.~T.~Crites} \affiliation{Kavli Institute for Cosmological Physics, University of Chicago, 5640 South Ellis Avenue, Chicago, IL, USA 60637} \affiliation{Department of Astronomy and Astrophysics, University of Chicago, 5640 South Ellis Avenue, Chicago, IL, USA 60637} \affiliation{Department of Physics, Cornell University, Ithaca, NY, USA} \affiliation{Department of Astronomy, Cornell University, Ithaca, NY, USA}
\author{T.~de~Haan} \affiliation{Department of Physics, University of California, Berkeley, CA, USA 94720} \affiliation{Physics Division, Lawrence Berkeley National Laboratory, Berkeley, CA, USA 94720} \affiliation{High Energy Accelerator Research Organization (KEK), Tsukuba, Ibaraki 305-0801, Japan}
\author{M.~A.~Dobbs} \affiliation{Department of Physics, McGill University, 3600 Rue University, Montreal, Quebec H3A 2T8, Canada} \affiliation{Canadian Institute for Advanced Research, CIFAR Program in Gravity and the Extreme Universe, Toronto, ON, M5G 1Z8, Canada}
\author[0000-0002-9962-2058]{D.~Dutcher} \affiliation{Joseph Henry Laboratories of Physics, Jadwin Hall, Princeton University, Princeton, NJ 08544, USA}
\author{W.~Everett} \affiliation{Department of Astrophysical and Planetary Sciences, University of Colorado, Boulder, CO, USA 80309}
\author{J.~Gallicchio} \affiliation{Kavli Institute for Cosmological Physics, University of Chicago, 5640 South Ellis Avenue, Chicago, IL, USA 60637} \affiliation{Harvey Mudd College, 301 Platt Blvd., Claremont, CA 91711}
\author{E.~M.~George} \affiliation{European Southern Observatory, Karl-Schwarzschild-Str. 2, 85748 Garching bei M\"{u}nchen, Germany} \affiliation{Department of Physics, University of California, Berkeley, CA, USA 94720}
\author[0000-0001-7652-9451]{N.~Gupta} \affiliation{CSIRO Space \& Astronomy, P.O. Box 1130, Bentley, WA 6102, Australia}
\author{N.~W.~Halverson} \affiliation{Department of Astrophysical and Planetary Sciences, University of Colorado, Boulder, CO, USA 80309} \affiliation{Department of Physics, University of Colorado, Boulder, CO, USA 80309}
\author[0000-0002-0463-6394]{G.~P.~Holder} \affiliation{Astronomy Department, University of Illinois at Urbana-Champaign, 1002 W. Green Street, Urbana, IL 61801, USA} \affiliation{Department of Physics, University of Illinois Urbana-Champaign, 1110 W. Green Street, Urbana, IL 61801, USA}
\author{W.~L.~Holzapfel} \affiliation{Department of Physics, University of California, Berkeley, CA, USA 94720}
\author{J.~D.~Hrubes} \affiliation{University of Chicago, 5640 South Ellis Avenue, Chicago, IL, USA 60637}
\author{N.~Huang} \affiliation{Department of Physics, University of California, Berkeley, CA, USA 94720}
\author{J.~Hubmayr} \affiliation{NIST Quantum Devices Group, 325 Broadway Mailcode 817.03, Boulder, CO, USA 80305}
\author{K.~D.~Irwin} \affiliation{SLAC National Accelerator Laboratory, 2575 Sand Hill Road, Menlo Park, CA 94025} \affiliation{Dept. of Physics, Stanford University, 382 Via Pueblo Mall, Stanford, CA 94305}
\author{L.~Knox} \affiliation{Department of Physics, University of California, One Shields Avenue, Davis, CA, USA 95616}
\author{A.~T.~Lee} \affiliation{Department of Physics, University of California, Berkeley, CA, USA 94720} \affiliation{Physics Division, Lawrence Berkeley National Laboratory, Berkeley, CA, USA 94720}
\author{D.~Li} \affiliation{NIST Quantum Devices Group, 325 Broadway Mailcode 817.03, Boulder, CO, USA 80305} \affiliation{SLAC National Accelerator Laboratory, 2575 Sand Hill Road, Menlo Park, CA 94025}
\author{A.~Lowitz} \affiliation{Department of Astronomy and Astrophysics, University of Chicago, 5640 South Ellis Avenue, Chicago, IL, USA 60637} \affiliation{Steward Observatory and Department of Astronomy, University of Arizona, 933 North Cherry Avenue, Tucson, AZ 85721, USA}
\author{J.~J.~McMahon} \affiliation{Kavli Institute for Cosmological Physics, University of Chicago, 5640 South Ellis Avenue, Chicago, IL, USA 60637} \affiliation{Department of Physics, University of Chicago, 5640 South Ellis Avenue, Chicago, IL, USA 60637} \affiliation{Department of Astronomy and Astrophysics, University of Chicago, 5640 South Ellis Avenue, Chicago, IL, USA 60637}
\author{J.~Montgomery} \affiliation{Department of Physics, McGill University, 3600 Rue University, Montreal, Quebec H3A 2T8, Canada}
\author{T.~Natoli} \affiliation{Department of Astronomy and Astrophysics, University of Chicago, 5640 South Ellis Avenue, Chicago, IL, USA 60637} \affiliation{Kavli Institute for Cosmological Physics, University of Chicago, 5640 South Ellis Avenue, Chicago, IL, USA 60637} \affiliation{Dunlap Institute for Astronomy \& Astrophysics, University of Toronto, 50 St George St, Toronto, ON, M5S 3H4, Canada}
\author{J.~P.~Nibarger} \affiliation{NIST Quantum Devices Group, 325 Broadway Mailcode 817.03, Boulder, CO, USA 80305}
\author[0000-0002-5254-243X]{G.~I.~Noble} \affiliation{Dunlap Institute for Astronomy \& Astrophysics, University of Toronto, 50 St George St, Toronto, ON, M5S 3H4, Canada} \affiliation{Department of Astronomy \& Astrophysics, University of Toronto, 50 St George St, Toronto, ON, M5S 3H4, Canada}
\author{V.~Novosad} \affiliation{Materials Sciences Division, Argonne National Laboratory, 9700 S. Cass Avenue, Argonne, IL, USA 60439}
\author{Y.~Omori} \affiliation{Department of Astronomy and Astrophysics, University of Chicago, 5640 South Ellis Avenue, Chicago, IL, USA 60637} \affiliation{Kavli Institute for Cosmological Physics, University of Chicago, 5640 South Ellis Avenue, Chicago, IL, USA 60637}
\author{S.~Padin} \affiliation{Kavli Institute for Cosmological Physics, University of Chicago, 5640 South Ellis Avenue, Chicago, IL, USA 60637} \affiliation{Department of Astronomy and Astrophysics, University of Chicago, 5640 South Ellis Avenue, Chicago, IL, USA 60637} \affiliation{California Institute of Technology, MS 367-17, 1216 E. California Blvd., Pasadena, CA, USA 91125}
\author{S.~Patil} \affiliation{School of Physics, University of Melbourne, Parkville, VIC 3010, Australia}
\author{C.~Pryke} \affiliation{School of Physics and Astronomy, University of Minnesota, 116 Church Street S.E. Minneapolis, MN, USA 55455}
\author{W.~Quan} \affiliation{High Energy Physics Division, Argonne National Laboratory, 9700 S. Cass Avenue, Argonne, IL, USA 60439} \affiliation{Department of Physics, University of Chicago, 5640 South Ellis Avenue, Chicago, IL, USA 60637} \affiliation{Kavli Institute for Cosmological Physics, University of Chicago, 5640 South Ellis Avenue, Chicago, IL, USA 60637}
\author[0000-0003-2226-9169]{C.~L.~Reichardt} \affiliation{School of Physics, University of Melbourne, Parkville, VIC 3010, Australia}
\author{J.~E.~Ruhl} \affiliation{Physics Department, Center for Education and Research in Cosmology and Astrophysics, Case Western Reserve University, Cleveland, OH, USA 44106}
\author{B.~R.~Saliwanchik} \affiliation{Physics Department, Center for Education and Research in Cosmology and Astrophysics, Case Western Reserve University, Cleveland, OH, USA 44106} \affiliation{Department of Physics, Yale University, P.O. Box 208120, New Haven, CT 06520-8120}
\author{K.~K.~Schaffer} \affiliation{Kavli Institute for Cosmological Physics, University of Chicago, 5640 South Ellis Avenue, Chicago, IL, USA 60637} \affiliation{Enrico Fermi Institute, University of Chicago, 5640 South Ellis Avenue, Chicago, IL, USA 60637} \affiliation{Liberal Arts Department, School of the Art Institute of Chicago, 112 S Michigan Ave, Chicago, IL, USA 60603}
\author{C.~Sievers} \affiliation{University of Chicago, 5640 South Ellis Avenue, Chicago, IL, USA 60637}
\author{G.~Smecher} \affiliation{Department of Physics, McGill University, 3600 Rue University, Montreal, Quebec H3A 2T8, Canada} \affiliation{Three-Speed Logic, Inc., Victoria, B.C., V8S 3Z5, Canada}
\author{A.~A.~Stark} \affiliation{Harvard-Smithsonian Center for Astrophysics, 60 Garden Street, Cambridge, MA, USA 02138}
\author{C.~Tucker} \affiliation{Cardiff University, Cardiff CF10 3XQ, United Kingdom}
\author{T.~Veach} \affiliation{Space Science and Engineering Division, Southwest Research Institute, San Antonio, TX 78238}
\author{J.~D.~Vieira} \affiliation{Astronomy Department, University of Illinois at Urbana-Champaign, 1002 W. Green Street, Urbana, IL 61801, USA} \affiliation{Department of Physics, University of Illinois Urbana-Champaign, 1110 W. Green Street, Urbana, IL 61801, USA}
\author{G.~Wang} \affiliation{High Energy Physics Division, Argonne National Laboratory, 9700 S. Cass Avenue, Argonne, IL, USA 60439}
\author[0000-0002-3157-0407]{N.~Whitehorn} \affiliation{Department of Physics and Astronomy, Michigan State University, 567 Wilson Road, East Lansing, MI 48824}
\author[0000-0001-5411-6920]{W.~L.~K.~Wu} \affiliation{Kavli Institute for Cosmological Physics, University of Chicago, 5640 South Ellis Avenue, Chicago, IL, USA 60637} \affiliation{SLAC National Accelerator Laboratory, 2575 Sand Hill Road, Menlo Park, CA 94025}
\author{V.~Yefremenko} \affiliation{High Energy Physics Division, Argonne National Laboratory, 9700 S. Cass Avenue, Argonne, IL, USA 60439}
\author{J.~A.~Zebrowski} \affiliation{Department of Physics, University of California, Berkeley, CA, USA 94720} \affiliation{Kavli Institute for Cosmological Physics, University of Chicago, 5640 South Ellis Avenue, Chicago, IL, USA 60637} \affiliation{Fermi National Accelerator Laboratory, MS209, P.O. Box 500, Batavia, IL 60510}

\collaboration{999}{(SPTpol collaboration)}



\begin{abstract}

Using the full four-year SPTpol 500 deg$^2$ dataset in both the 95~GHz and 150~GHz frequency bands, we present measurements of the temperature and $E$-mode polarization of the cosmic microwave background (CMB), as well as the $E$-mode polarization auto-power spectrum ($EE$) and temperature-$E$-mode cross-power spectrum ($TE$) in the angular multipole range $50<\ell<8000$. We find the SPTpol dataset to be self-consistent, passing several internal consistency tests based on maps, frequency bands, bandpowers, and cosmological parameters. The full SPTpol dataset is well-fit by the $\Lambda CDM$ model, for which we find $H_0=70.48\pm2.16$~km~s$^{-1}$~Mpc$^{-1}$ and $\Omega_m=0.271\pm0.026$, when using only the SPTpol data and a \textit{Planck}-based prior on the optical depth to reionization. The $\Lambda CDM$ parameter constraints are consistent across the 95~GHz-only, 150~GHz-only, $TE$-only, and $EE$-only data splits. Between the $\ell<1000$ and $\ell>1000$ data splits, the $\Lambda CDM$ parameter constraints are borderline consistent at the $\sim2\sigma$ level. This consistency improves when including a parameter $A_L$, the degree of lensing of the CMB inferred from the smearing of acoustic peaks. When marginalized over $A_L$, the $\Lambda CDM$ parameter constraints from SPTpol are consistent with those from \textit{Planck}. The power spectra presented here are the most sensitive measurements of the lensed CMB damping tail to date for roughly $\ell > 1700$ in $TE$ and $\ell > 2000$ in $EE$.
\end{abstract}



\section{Introduction} \label{sec:intro}

Measurements of the temperature and polarization of the cosmic microwave background (CMB) provide valuable information about cosmology \citep[e.g.,][]{hu02b,galli14}. Their angular power spectra are well-fit by the $\Lambda CDM$ model, and currently the best constraints for most of the $\Lambda CDM$ parameters come from CMB experiments, in particular the \textit{Planck} satellite \citep{planck18-6}, but also ground-based experiments such as the Atacama Cosmology Telescope \citep[ACT, e.g.,][]{aiola20,choi20} and the South Pole Telescope \citep[SPT, e.g.,][]{balkenhol23,dutcher21}.
In recent years, tensions have emerged between some of these parameter constraints and those measured from the late-time universe \citep[e.g,][]{verde23,heymans20}, and there have even been low-level ($\sim 2\sigma$) hints of disagreements within different CMB data sets or subsets of individual-experiment data \citep{henning18,dutcher21,planck18-6}. With an eye toward sharpening or disfavoring one of these observed trends, in this work we present a deeper investigation into the internal consistency of the SPTpol 500~deg$^2$ dataset, adding more data volume and multifrequency observations to the analysis in \citet[][hereafter H18]{henning18}.

The SPT \citep{carlstrom11}, with its arcminute-scale resolution and polarization-sensitive detectors, probes an important space of the CMB power spectra. Specifically, the SPTpol 500~deg$^2$ measurements in this work cover a wide range of angular modes, from the first peak to the damping tail, in both temperature and polarization. SPTpol \citep{henning12,sayre12} was the 2nd generation receiver installed on SPT, and it was composed of 1536 transition edge sensor detectors with observing bands centered at 95 or 150~GHz. This work is an extension to H18, where we perform $TE$/$EE$ power spectrum analysis in a similar fashion.
$TT$ is omitted in these analyses to avoid the need to model complex foregrounds, such as the cosmic infrared background and the thermal Sunyaev-Zel'dovich effect, which are expected to be largely unpolarized. 
H18 only analyzed data from the 150~GHz detectors, and only from the first three years of 500~deg$^2$ observations; in this work, we analyze the full four-year dataset, and we use data from both the 95~GHz and the 150~GHz frequency bands. In H18, the data used to report the entire multipole range of bandpowers were processed identically, but in this work, we process the data in two different ways into two datasets: a low-$\ell$ focused one and a high-$\ell$ focused one.
All these extensions combined allow us to make the most sensitive measurements of the lensed CMB damping tail to date for roughly $\ell > 1700$ in $TE$ and $\ell > 2000$ in $EE$. (Recently published estimates of the primordial, \textit{unlensed} $EE$ power spectrum from SPT-3G data in \citet{ge24} are more sensitive than the $EE$ spectrum presented here for $400 < \ell < 3500$, the entire $\ell$ range covered in that work.) We present our 500~deg$^2$ temperature and $E$-mode polarization maps, as well as our $TE$ and $EE$ angular power spectra in the multipole range $50<\ell<8000$. 

We provide an update to the mild tension between low-$\ell$ and high-$\ell$ cosmological parameter preferences found in H18. H18 had unresolved curiosities including internal tensions and poor fits to $\Lambda CDM$, some of which could be due to unmodeled systematics. In this work we cast a wider net searching for systematic effects, and we make several improvements including stringent cuts, extensive null tests, and updated beam measurements. We follow up on the aforementioned curiosities with these improved measurements of the SPTpol 500~deg$^2$ $TE$/$EE$ power spectra.

This paper is structured as follows. We describe the SPTpol observations and time-ordered data in Section~\ref{sec:data}. From these data, we make maps and present them in Section~\ref{sec:maps}. From these maps, we make power spectra with methods described in Section~\ref{sec:powspec}, and we present the binned power spectra (bandpowers) in Section~\ref{sec:bandpowers}. We fit these bandpowers to cosmological models with the methods described in Section~\ref{sec:likelihood}, and we present the parameter constraints in Section~\ref{sec:constraints}. We discuss the conclusions in Section~\ref{sec:conclusion}.

\section{Observations and Data Reduction} \label{sec:data}


The SPTpol instrument, the 500~deg$^2$ survey field, and the scan strategies are the same as H18; please refer to that work for more details. H18 only analyzed data from the 150~GHz detectors and only from the first three years of observations; in this work, we analyze the full four-year dataset and use data from both the 95~GHz and the 150~GHz frequency bands.

\subsection{Observations} \label{subsec:obs}

An observation is one complete mapping of the SPTpol 500 deg$^2$ survey field, in which we make $\sim100$ pairs of constant-elevation right-going and left-going scans, followed by a discrete step in elevation. Each single observation lasts approximately two hours and covers the entire declination range of the field. 
Observations from April 2013 to May 2014 were made using the ``lead-trail'' scan strategy, where the 500~deg$^2$ field was split into two halves in right ascension, with one half observed after the other in the same azimuth-elevation range. Observations from May 2014 to Sep 2016 were made using the ``full-field'' scan strategy, where we scanned across the full range of right ascension of the 500~deg$^2$ field. For each observation, we produce a 95~GHz map and a 150~GHz map, however one or both of them can fail during the time stream processing described in Section~\ref{subsec:tod}. Unlike H18, we do not perform observation cuts based on low-$\ell$ noise, but we remove $\sim200$ observations based on jackknife null tests described in Section~\ref{subsec:jackknife}. The final number of 95~GHz observations used in this dataset is 1481 lead-trail \& 3368 full-field, and the number of 150~GHz observations used in this dataset is 1481 lead-trail \& 3387 full-field.

\subsection{Time Stream Processing} \label{subsec:tod}

We record the time-ordered data (time streams) of each detector, and they are subject to several processes before being combined into maps. As in H18, spectral lines related to the pulse tube coolers are notch-filtered; additionally, in this work the 95~GHz time streams show a strong 1 Hz noise line, so that line is also notch-filtered. To further improve signal-to-noise at high $\ell$ over H18, we then analyze two copies of the time streams; we apply a high-$\ell$ focused set of filters on one copy, and apply a low-$\ell$ focused set of filters on the other copy.



For the high-$\ell$ focused time streams, we subtract a 5th-order (or 9th-order) Legendre polynomial on a scan-by-scan basis for lead-trail (or full-field) observations. To prevent noise at low $\ell_x$ from mixing into high-$\ell$ modes (where $\ell_x$ is along the scan direction), we apply a high-pass filter at $\ell_x=300$, and a common-mode filter that removes the average of all time streams in the same frequency band. For anti-aliasing, we apply a low-pass filter at $\ell_x=20{,}000$.

For the low-$\ell$ focused time streams, we first downsample the data by a factor of 2, and then subtract the same 5th-order (or 9th-order) Legendre polynomial for lead-trail (or full-field) observations. This step effectively equals a high-pass filter at $\ell_x \sim 50$, and we apply no further high-pass filters. For anti-aliasing, we apply a low-pass filter at $\ell_x=3{,}200$.

Both copies of the time streams go through the same cross-talk removal, detector data cuts, and pre-map calibration as H18, except we make one additional detector cut on the high-$\ell$ focused time stream data due to a noise line. Detectors with anomalously high noise in the 8--11 Hz frequency band are cut on a per-observation basis. 

\section{Maps} \label{sec:maps}

As in H18, we bin the time streams into $T,Q,U$ map pixels under the oblique Lambert azimuthal equal-area projection.
For the high-$\ell$ focused time streams, we use square 0.$5^\prime$~pixels, and for the low-$\ell$ focused time streams, we use square $3^\prime$~pixels to speed up computations. From these low-$\ell$ focused maps only, we remove a smooth scan-synchronous structure. To correct for monopole $T$-to-$P$ leakage (see Section~\ref{subsec:TtoP}), we subtract a scaled copy of the $T$ map from the $Q$ and $U$ maps, and then we construct the $E$-mode polarization map from $Q$ and $U$ in the same way as H18 \citep{zaldarriaga01}.

\begin{figure*}[htb]
\plotone{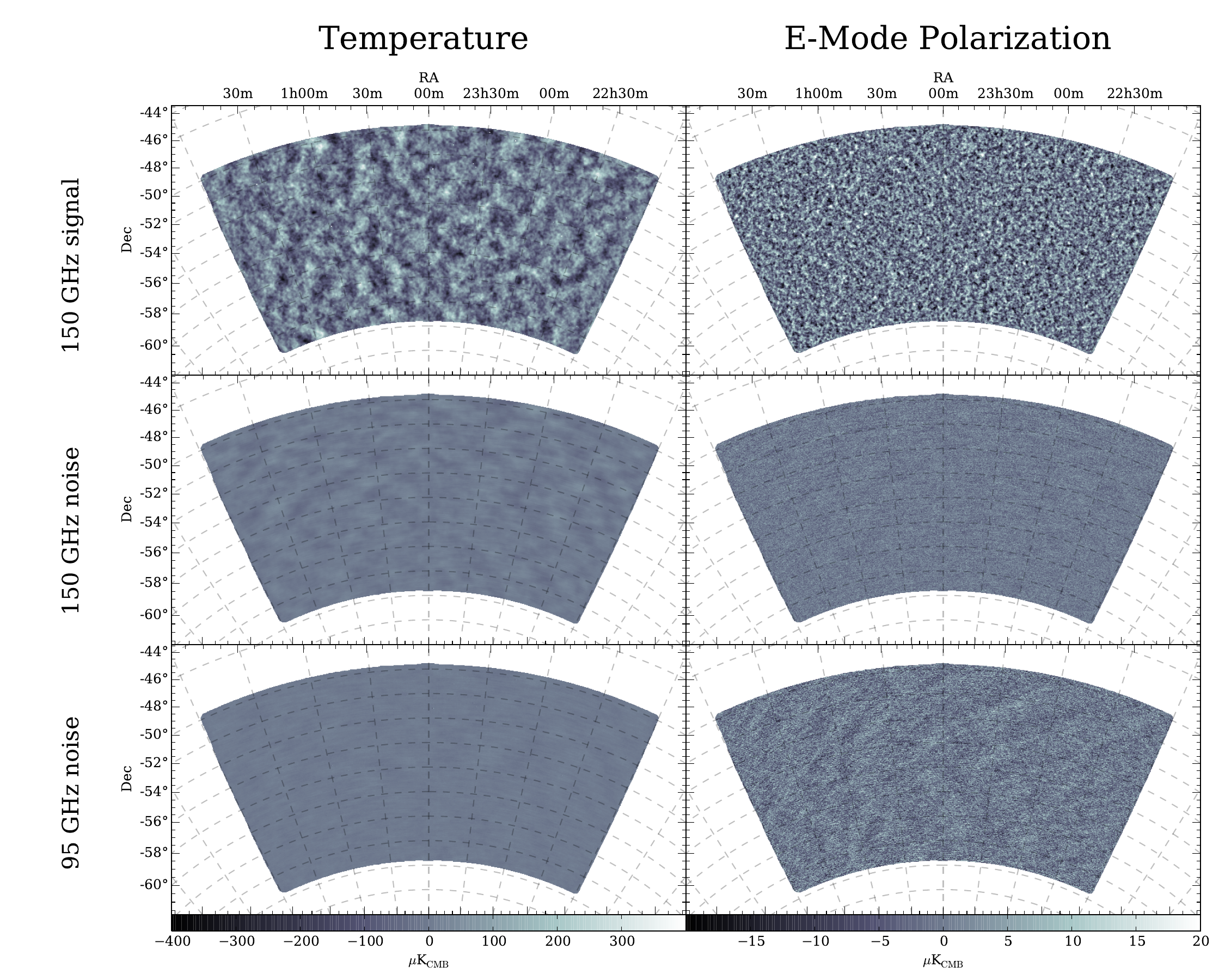}
\caption{SPTpol 500~deg$^2$ low-$\ell$ focused signal and noise maps, for temperature and $E$-mode polarization, for 150~GHz and 95~GHz. The 95~GHz signal maps are omitted because they look similar to the 150~GHz signal maps. The noise maps are made with the coadd of left-going scans minus the coadd of right-going scans, then divided by 2.
\label{fig:map3x2}}
\end{figure*}

\begin{figure*}[htb]
\plotone{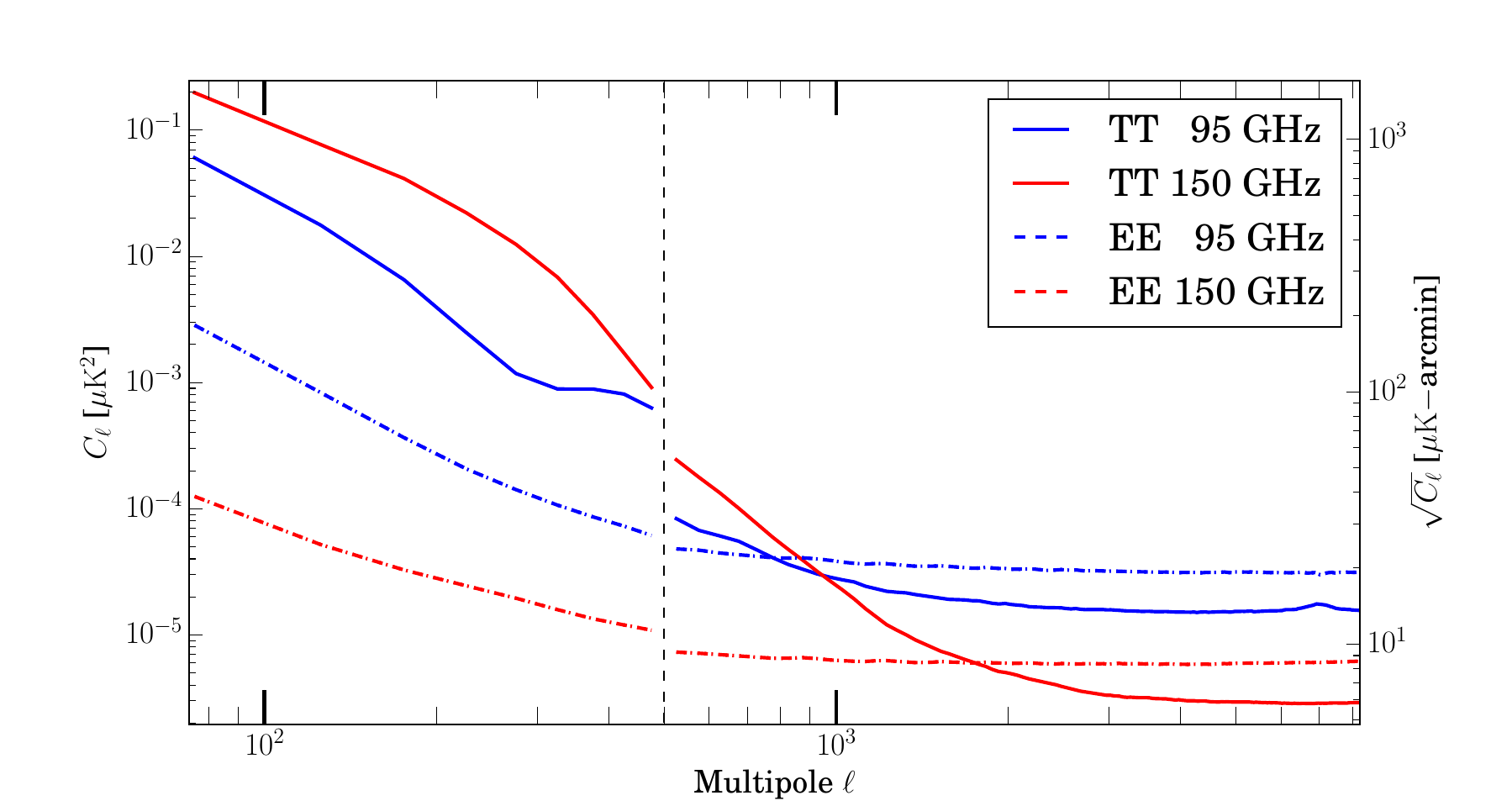}
\caption{SPTpol 500~deg$^2$ noise spectra. For $\ell<500$, we use the low-$\ell$ focused dataset; for $\ell>500$, we use the high-$\ell$ focused dataset, and the dashed line at $\ell=500$ shows this stitch. These are unbiased spectra but multiplied by the beam function again ($B_\ell^2$) to show the white noise level. Units on both axes are provided as a handy reference for conversion between $\mu K^2$ and $\mu K$-arcmin.
\label{fig:noise}}
\end{figure*}

Figure~\ref{fig:map3x2} shows the coadd (inverse-variance-weighted average) of the 150~GHz low-$\ell$ focused maps, and the 150~GHz and 95~GHz noise maps, for temperature and $E$-mode polarization. The 95~GHz signal maps are omitted because they look similar to the 150~GHz signal maps.
Figure~\ref{fig:noise} shows the noise spectra of these 95~GHz and 150~GHz temperature and $E$-mode polarization maps. The white noise level in the multipole range $5000<\ell\leq6000$ is 5.9~$\mu$K-arcmin for the 150~GHz temperature map, 8.4~$\mu$K-arcmin for the 150~GHz $E$-mode map, 13.5~$\mu$K-arcmin for the 95~GHz temperature map, and 19.2~$\mu$K-arcmin for the 95~GHz $E$-mode map.

Similar to H18, we make 50 partial coadds (map bundles) for lead-trail and full-field separately, where each bundle is constructed to have 1/50th of the total $T$ map weights. Each lead-trail bundle is then coadded with a full-field bundle in chronological order, and these 50 ``lead-trail plus full-field'' map bundles will be used in Section~\ref{subsec:xspec} to calculate the angular power spectra. In total, we have 50 map bundles for 150~GHz and 50 for 95~GHz, where each bundle has a $T$ map and an $E$ map. Lastly, we make noise map realizations by randomly separating all the $\sim4800$ maps into two subsets, coadding each subset, and subtracting these two half-depth coadded maps. We do all of this for the low-$\ell$ focused maps and the high-$\ell$ focused maps separately.

\section{Power Spectrum} \label{sec:powspec}

\subsection{Pseudo Cross-Spectra} \label{subsec:xspec}

We compute angular power spectra with the same pseudo-spectrum method as H18 \citep{hivon02}, using cross-spectra to avoid noise bias \citep{polenta05,tristram05,lueker10}. From among our 50 map bundles in 150~GHz and 50 bundles in 95~GHz, we compute cross-spectra of every possible pair, never crossing the same bundle with itself. We compute all seven possible $TE$ or $EE$ cross-spectra ($95T\times95E$, $95T\times150E$, $150T\times95E$, $150T\times150E$, $95E\times95E$, $95E\times150E$, $150E\times150E$), and take the average within each type. We do these for the low-$\ell$ focused maps and the high-$\ell$ focused maps separately.

These cross-spectra are computed at a native $\ell$ resolution of $\Delta\ell=5$. We denote them as $\widetilde{C}_\lambda$, where $\lambda$ represents the resolution of $\Delta\ell=5$, and the tilde above indicates that these are pseudo-spectra. The pseudo-spectrum can be represented as the result of a biasing kernel $K$ acting on the unbiased power spectrum \citep{hivon02}; equivalently, the unbiased power spectrum equals the inverse of the biasing kernel acting on the pseudo-spectrum, and we perform this inverse after binning to a resolution of $\Delta\ell=50$. In Einstein summation notation,
\begin{align}
\widetilde{C}_\beta &= K_{\beta\beta'}C_{\beta'} \nonumber\\
\Rightarrow C_{\beta} &= K_{\beta\beta'}^{-1}\widetilde{C}_{\beta'} \nonumber\\
          &= K_{\beta\beta'}^{-1}P_{\beta'\lambda}\widetilde{C}_{\lambda} \label{eq:MASTER} \,,
\end{align}
where $\beta$ denotes a resolution of $\Delta\ell=50$, and $P_{\beta\lambda}$ is defined as a binning operator that bins bandpowers from a resolution of $\lambda$ into a resolution of $\beta$. The biasing kernel $K$ consists of the mode-coupling matrix $M$ due to the apodization mask, the filter transfer function $F$, and the beam function $B$:
\begin{align}
K_{\lambda\lambda'} &= M_{\lambda\lambda'}F_{\lambda'}B_{\lambda'}^2 \label{eq:Kll} \,.
\end{align}
In the next three subsections, we will discuss these three components in more detail. They are computed at a native resolution of $\Delta\ell=5$, and then the biasing kernel is binned to $\Delta\ell=50$:
\begin{align}
K_{\beta\beta'} &= P_{\beta\lambda}K_{\lambda\lambda'}Q_{\lambda'\beta'} \label{eq:Kbb} \,,
\end{align}
where $Q_{\lambda\beta}$ is the reciprocal to the binning operator $P_{\beta\lambda}$. After obtaining the unbiased cross-spectra $C_\beta$, we convert them to $D_\beta$ using the operator $S_{\beta\beta}=\ell^{(\beta)}(\ell^{(\beta)}+1)/2\pi$ where $\ell^{(\beta)}$ is the $\ell$ center of bin $\beta$. Finally, we coarse-bin them into increasingly wider $\ell$ bins starting at $\ell\geq2000$, in order to reduce the total number of bandpowers and simplify numerical computations.
\begin{align}
D_b = P_{b\beta}S_{\beta\beta}C_{\beta} \label{eq:cl2dl} \,.
\end{align}

These final $\ell$ bins and $D_b$ are shown in Section~\ref{sec:bandpowers}, and they are used in Section~\ref{sec:likelihood} for fitting cosmological parameters. In addition, we also compute unbiased noise spectra from the noise map realizations mentioned in the previous section, and they will be used in the construction of the bandpower covariance matrix in Section~\ref{subsec:covmat}.

\subsection{Mask and Mode-Coupling} \label{subsec:apod}

We make our apodization and point source mask in the same way as H18, where all point sources with unpolarized flux $>50$~mJy at 95 or 150~GHz are masked. We also compute the mode-coupling matrix $M_{\lambda\lambda'}$ analytically, in the same way as H18. The mode-coupling matrices in this work and in H18 both exhibit band-diagonal structures, where elements the same distance away from the diagonal are approximately equal, therefore we average them together during the construction of the bandpower covariance matrix in Section~\ref{subsec:covmat}. We do these for the low-$\ell$ focused maps and the high-$\ell$ focused maps separately.


\subsection{Transfer Function} \label{subsec:xfer}

The filter transfer function $F$ accounts for the effects of time stream processing described in Section~\ref{subsec:tod}. As in H18, we solve for it using simulated skies and ``mock-observations.'' We make 226 realizations of the sky from a given CMB power spectrum $C_\ell$, where $C_\ell$ is the best-fit theory to the \textit{Planck} \verb|base_plikHM_TT_lowTEB_lensing| dataset \citep{planck18-6}. Next, we add a realization of the foreground power to each sky realization. The foreground power spectrum is modeled as follows:
\begin{align*}
\textrm{foreground } D_\ell^{TT} &= \left(A_\mathrm{src}^\mathrm{dusty}+A_\mathrm{src}^\mathrm{radio}\right) \left(\frac{\ell}{3000}\right)^2 \\
&\enspace + A_\mathrm{CIB} \left(\frac{\ell}{3000}\right)^{0.8} + A_\mathrm{tSZ}\cdot\mathrm{template}\\
\textrm{foreground } D_\ell^{EE} &= \left(A_\mathrm{src}^\mathrm{dusty}p^\mathrm{dusty} + A_\mathrm{src}^\mathrm{radio}p^\mathrm{radio}\right) \left(\frac{\ell}{3000}\right)^2 \\
&\enspace + A_\mathrm{dust}^{EE} \left(\frac{\ell}{80}\right)^{-0.42} \,,
\end{align*}
where the mean-squared polarization fraction for dusty point sources $p^\mathrm{dusty}=0.0004$, for radio point sources $p^\mathrm{radio}=0.0014$, and the tSZ model template is taken from \citet{shaw10}. For 150~GHz, the amplitudes in $\mu K^2$ are: \{$A_\mathrm{src}^\mathrm{dusty}=9$, $A_\mathrm{src}^\mathrm{radio}=10$, $A_\mathrm{CIB}=3.46$, $A_\mathrm{tSZ}=4$, $A_\mathrm{dust}^{EE}=0.0236$\}. For 95~GHz, the amplitudes in $\mu K^2$ are: \{$A_\mathrm{src}^\mathrm{dusty}=1.5$, $A_\mathrm{src}^\mathrm{radio}=50$, $A_\mathrm{CIB}=0.56$, $A_\mathrm{tSZ}=12$, $A_\mathrm{dust}^{EE}=0.00338$\}. 

We convolve these sky realizations with the beam function $B_\ell$, which is different for 95~GHz and 150~GHz. Next, we mock-observe these sky realizations by scanning through them at the same telescope pointings as recorded in each of our $\sim4800$ observations. Time stream processing, map-making and bundling are also done in the same way as real data, and the end product is 226 simulated datasets: We have all the low-$\ell$ focused and high-$\ell$ focused map bundles, for 95 or 150~GHz, for each sky realization.

For each of the 226 simulated datasets, we compute $95T\times95T$, $150T\times150T$, $95E\times95E$, $150E\times150E$, and $95E\times150E$.
These are pseudo-spectra, and the input theory $C_\ell^{TT}$ and $C_\ell^{EE}$ are known, so we solve for the filter transfer functions iteratively as in H18 \citep{hivon02}. For the five types of $TT$ or $EE$ spectra above, we compute their transfer functions one at a time by plugging in the average $\widetilde{C}_\lambda$ of 226 simulations and $C_{\lambda,th}$ (again, $\lambda$ denotes a resolution of $\Delta\ell=5$):
\begin{align*}
F_{\lambda,1} &= \frac{\langle\widetilde{C}_\lambda^\mathrm{\:sim}\rangle}{w_2C_{\lambda,th}B_\lambda^2}\\
F_{\lambda,i+1} &= F_{\lambda,i} + \frac{\langle\widetilde{C}_\lambda^\mathrm{\:sim}\rangle - M_{\lambda\lambda'}F_{\lambda',i}C_{\lambda',th}B_{\lambda'}^2}{w_2C_{\lambda,th}B_\lambda^2} \,,
\end{align*}
where $w_2$ denotes the second moment of the apodization mask, and the index $i$ runs from 1 to 2 because the computation converges after two iterations, as was the case in H18. This method could be numerically unstable for the $TE$ spectra due to their zero-crossings, therefore we define the $TE$ transfer functions as the geometric mean of the corresponding $TT$ and $EE$ transfer functions:
\begin{align*}
F_\lambda^{95T\times95E} &= \sqrt{F_\lambda^{95T\times95T}\cdot F_\lambda^{95E\times95E}}\\
F_\lambda^{95T\times150E} &= \sqrt{F_\lambda^{95T\times95T}\cdot F_\lambda^{150E\times150E}}\\
F_\lambda^{150T\times95E} &= \sqrt{F_\lambda^{150T\times150T}\cdot F_\lambda^{95E\times95E}}\\
F_\lambda^{150T\times150E} &= \sqrt{F_\lambda^{150T\times150T}\cdot F_\lambda^{150E\times150E}} \,.
\end{align*}

These transfer functions $F_\lambda$ are used in unbiasing our bandpowers as described in Section~\ref{subsec:xspec}. For later use, we also unbias these simulated spectra $\widetilde{C}_\lambda$ using the same method. In addition, we repeat the above simulation process but with an ``alternate'' cosmology instead of \textit{Planck}, and these simulated spectra will be used in the pipeline consistency tests (Section~\ref{subsec:pipelineTest}). We do all of this for the low-$\ell$ focused maps and the high-$\ell$ focused maps separately.

\subsection{Beam and Calibration} \label{subsec:beam}

We use Venus observations to estimate the beam function $B_\ell$ of our instrument. For 150~GHz, we use the same observations as H18.
For 95~GHz, we also use Venus observations to estimate the beam function, but we note that there are map artifacts in the 95~GHz Venus maps at angular scales corresponding to $\ell \gg 8000.$ We compare the 95~GHz $B_\ell$ measured on Venus to the $B_\ell$ measured on Mars which does not contain those artifacts, and find that the two are consistent at $\ell<8000$. Since we only report bandpowers at $\ell < 8000$, and since the Venus beam has better sensitivity than the Mars beam at low-$\ell$, we decide to use the Venus observations for the 95~GHz beam as well.

The instantaneous pointing of the SPT is stable on hour timescales but drifts on day timescales owing to changes in the thermal environment. This pointing ``jitter'' is negligible for planet observations, but for averages of many CMB field scans, it causes the effective beam size to increase.
We estimate this by fitting 2D Gaussians to the brightest point sources in the 500~deg$^2$ field. For 150(95)~GHz, the final size of the Gaussian beam is $1.22^\prime(1.90^\prime)$~FWHM. We convolve each Venus map with a 2D Gaussian whose FWHM is the quadrature difference between $1.22^\prime$ (or $1.90^\prime$) and the FWHM of Venus in that map. For 95~GHz and 150~GHz separately, we take the cross-spectra between these maps, and their average is our $B_\ell^2$.

\begin{figure*}[htb]
\epsscale{0.8}
\plotone{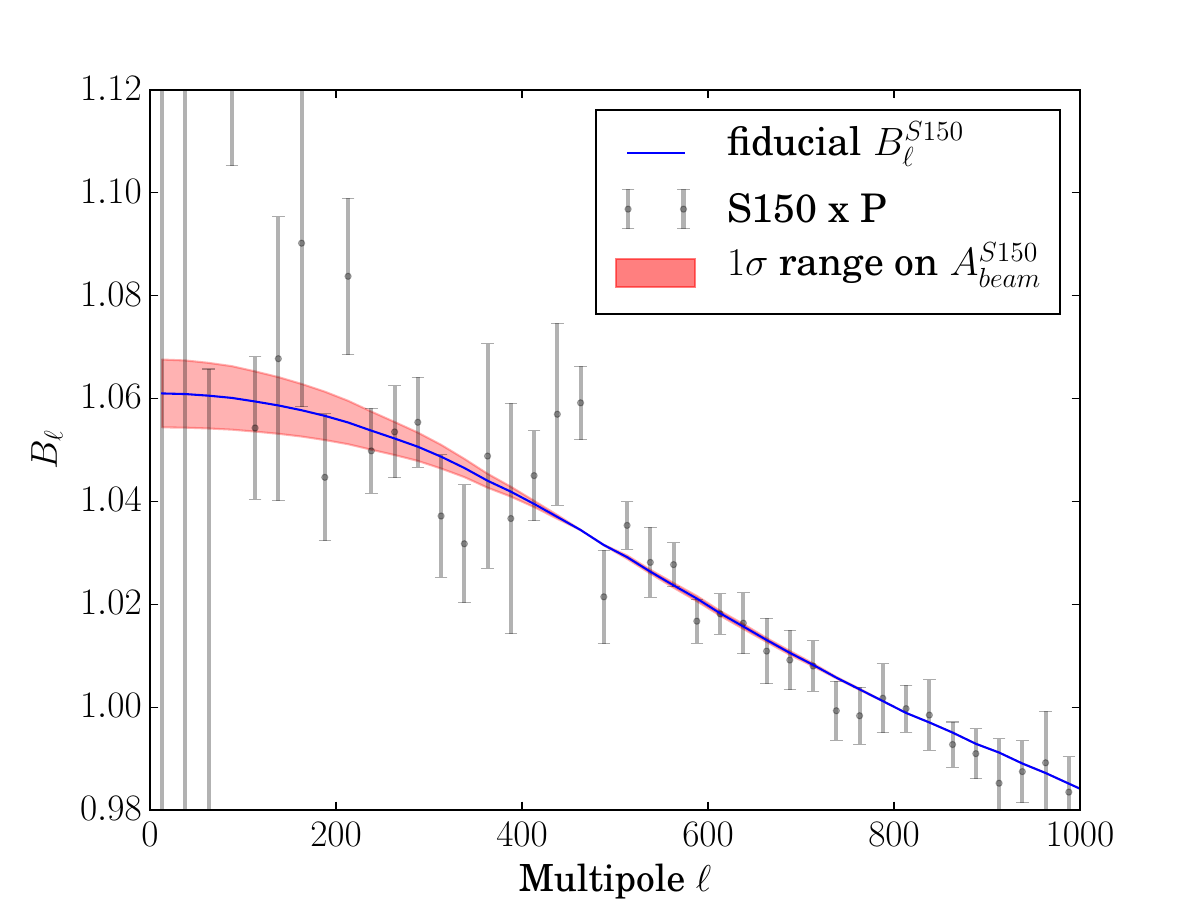}
\caption{The fiducial SPTpol 150~GHz beam (blue), and on top of it, the range of $\Delta B_\ell^{S150}$ that corresponds to 1-$\sigma$ uncertainties in the $A_\textrm{beam}^\textrm{S150}$ parameter (red region). Gray data points with error bars represent the right hand side of Eq.~\ref{eq:lowell_beam} with uncertainties.
\label{fig:lowell_beam}}
\end{figure*}

Due to the insufficient size of these Venus maps, we discovered that $B_\ell$ at $\ell<800$ is not precise, and can vary greatly depending on our Venus mapmaking options. This low-$\ell$ beam uncertainty can be well described by a one-parameter model $B_\ell \rightarrow B_\ell + A_\textrm{beam}\cdot \Delta B_\ell$ where $\Delta B_\ell$ is a known vector describing the typical shape of this variation, and $A_\textrm{beam}$ is a free parameter of its amplitude. For the case of 150~GHz, we fit for $A_\textrm{beam}$ and the temperature calibration factor $T_\textrm{cal}$ at the same time against \textit{Planck}. As in H18, we mock-observe the \textit{Planck} 143~GHz temperature map in the same way as our observations, and take its cross-spectrum with the SPTpol 150~GHz temperature map:

\begin{align} \label{eq:lowell_beam}
& T_\textrm{cal}^\textrm{S150} \left(B_\ell^\textrm{S150} + A_\textrm{beam}^\textrm{S150}\cdot \Delta B_\ell^\textrm{S150}\right) \nonumber\\
=\enspace& \sqrt{F_\ell^P}B_\ell^P
\frac{\left\langle\textrm{Re}\left[\textrm{fmap}_1^{*,\textrm{S150}}\cdot \textrm{fmap}_2^\textrm{S150}\right]\right\rangle_\ell}{\left\langle\textrm{Re}\left[\textrm{fmap}^{*,P}\cdot \textrm{fmap}^\textrm{S150}\right]\right\rangle_\ell} \,,
\end{align}
where the superscript P denotes \textit{Planck} 143~GHz, S150 denotes SPTpol 150~GHz, fmap denotes the temperature map in Fourier space, and the subscripts in fmap$_1$ or fmap$_2$ denote two independent half-depth maps of SPTpol 150~GHz. The transfer function $\sqrt{F_\ell^P}$ denotes the pixel window function that corresponds to the size of the \textit{Planck} map pixels.
We compute the right hand side with uncertainties, then perform a least-squares fit over the range $100 < \ell < 1000$. Under the best-fit $T_\textrm{cal}^\textrm{S150}$ and $A_\textrm{beam}^\textrm{S150}$ values, Figure~\ref{fig:lowell_beam} shows what we now define as the ``fiducial'' $B_\ell^\textrm{S150}$, and for illustration purposes, we also show the range of $\Delta B_\ell^\textrm{S150}$ that corresponds to 1-$\sigma$ uncertainties in the $A_\textrm{beam}^\textrm{S150}$ parameter. For the case of 95~GHz, we similarly fit for $A_\textrm{beam}$ and $T_\textrm{cal}$ at the same time against 150~GHz:

\begin{align*}
& T_\textrm{cal}^\textrm{S95} \left(B_\ell^\textrm{S95} + A_\textrm{beam}^\textrm{S95}\cdot \Delta B_\ell^\textrm{S95}\right) \\
=\enspace& T_\textrm{cal}^\textrm{S150}\cdot \sqrt{\frac{F_\ell^\mathrm{S150}}{F_\ell^\mathrm{S95}}}\cdot \textrm{fiducial }B_\ell^\textrm{S150} \\
\cdot &\; \frac{\left\langle\textrm{Re}\left[\textrm{fmap}_i^{*,\textrm{S95}}\cdot \textrm{fmap}_{i+1}^\textrm{S150}\right]\right\rangle_\ell}{\left\langle\textrm{Re}\left[\textrm{fmap}_{1,i}^{*,\textrm{S150}}\cdot \textrm{fmap}_{2,i}^\textrm{S150}\right]\right\rangle_\ell} \,,
\end{align*}
where S95 denotes SPTpol 95~GHz, the subscript $i$ denotes the $i$-th map bundle, and fmap$_{1,i}^\textrm{S150}$ or fmap$_{2,i}^\textrm{S150}$ denote two independent half-depth maps of the $i$-th 150~GHz map bundle. We allow $i$ to run through all 50 bundles and take the average. Similar to the above, we define the fiducial $B_\ell^\textrm{S95}$ using the best-fit values of $T_\textrm{cal}^\textrm{S95}$ and $A_\textrm{beam}^\textrm{S95}$. For the polarization calibration factor $P_\textrm{cal}$, we repeat the above process using $E$-mode polarization maps instead of temperature maps. We adopt the value $P_\textrm{cal}^{S150} = 1.06$ found in H18, and in this work we find $P_\textrm{cal}^{S95} = 1.043$. When performing cosmology fits, we still allow these $T_\textrm{cal},P_\textrm{cal},$ and beam uncertainty parameters to float as nuisance parameters, with their 1-$\sigma$ uncertainty ranges as priors (Section~\ref{subsec:nuisance}).

\subsection{Bandpower Window Functions} \label{subsec:bpwf}

To compare theory spectra $C_l^{th}$ with our bandpowers $D_b$, we need to find the bandpower window functions that convert $C_l^{th}$ to the equivalent $D_b^{th}$.
We begin by defining the binned pseudo-spectrum $\widetilde{C}_\lambda^{th}$ as the theory $C_l^{th}$ acted on by the biasing kernel and the initial binning operator,
\begin{align*}
\widetilde{C}_{\lambda}^{th} &= K_{\lambda\lambda'}P_{\lambda'l}C_l^{th} \,.
\end{align*}
Plugging this into Eq.~\ref{eq:MASTER} yields the expression for the unbiased theory spectrum at intermediate binning,
\begin{align*}
C_{\beta}^{th} &= K_{\beta\beta'}^{-1}P_{\beta'\lambda}K_{\lambda\lambda'}P_{\lambda'l}C_l^{th} \,.
\end{align*}
And Eq.~\ref{eq:cl2dl} yields the expression for the unbiased theory spectrum at final binning, including the $\ell(\ell+1)/2\pi$ weighting,
\begin{align*}
\textrm{final } D_b^{th} =\enspace& P_{b\beta}S_{\beta\beta}C_{\beta}^{th} \\
=\enspace& \left(P_{b\beta}S_{\beta\beta}K_{\beta\beta'}^{-1}P_{\beta'\lambda}K_{\lambda\lambda'}P_{\lambda'l}\right)\;C_l^{th} \,.
\end{align*}

The quantity in parentheses above is our bandpower window function. $K_{\lambda\lambda'}$ and $K_{\beta\beta'}$ are defined in Eq.~\ref{eq:Kll} and Eq.~\ref{eq:Kbb}.

\subsection{T-to-P Leakage} \label{subsec:TtoP}

Through various systematic effects, it is possible for the temperature measurements to leak into polarization measurements, and the polarization maps need to be corrected for this effect.
For the monopole leakage, we subtract a scaled copy of the $T$ map from the $Q$ and $U$ maps: $Q\rightarrow Q-\epsilon^Q T$ and $U\rightarrow U-\epsilon^U T$, where the coefficients $\epsilon^Q$ and $\epsilon^U$ are estimated from the ratio of half-dataset cross-correlated maps. For the low-$\ell$ focused dataset, we estimate $\epsilon^Q$ to be $0.016$($0.025$) for 150(95)~GHz, and $\epsilon^U$ to be $0.009$($-0.022$) for 150(95)~GHz, with an uncertainty of $\pm0.001$. For the high-$\ell$ focused dataset, we estimate $\epsilon^Q$ to be $0.020$($0.032$) for 150(95)~GHz, and $\epsilon^U$ to be $0.007$($-0.022$) for 150(95)~GHz, with an uncertainty of $\pm0.002$.

We also test for $T$-to-$P$ leakage beyond a monopole, which we refer to as the ``leakage beam.'' The leakage beam for 95~GHz is negligible, and the leakage beam for 150~GHz can be well explained by differential beam ellipticity, the effect of which in the power spectrum we parameterize as:
\begin{align*}
C_{\ell,\mathrm{corrected}}^{TE} &= C_{\ell,\mathrm{observed}}^{TE} - \ell^2 G_\ell^{TE} C_\ell^{TT} \,,\\
C_{\ell,\mathrm{corrected}}^{EE} &= C_{\ell,\mathrm{observed}}^{EE} - \ell^4 G_\ell^{EE} C_\ell^{TT} \,.
\end{align*}
We find best-fit values of $G_\ell^{TE} = 7.38\times10^{-10}$ and $G_\ell^{EE} = 6.40\times10^{-19}$.

As in \citet{dutcher21}, the common-mode filter introduces a bias in the $TE$ spectra, also the $TE$ transfer function is not designed to perfectly recover the simulated $TE$ spectra. We define this ``$TE$ bias'' as the difference between unbiased simulated bandpowers and theory bandpowers with window function:
\begin{align*}
TE\textrm{bias}^{i\times j} &= C_{\beta,sim}^{i\times j} - C_{\beta,th}^{i\times j}\\
&= K_{\beta\beta'}^{-1}P_{\beta'\lambda}\widetilde{C}_{\lambda,sim}^{i\times j} - K_{\beta\beta'}^{-1}P_{\beta'\lambda}K_{\lambda\lambda'}P_{\lambda'l}C_{l,th}^{i\times j}\\
&= K_{\beta\beta'}^{-1}P_{\beta'\lambda} \left(\widetilde{C}_{\lambda,sim}^{i\times j} - K_{\lambda\lambda'}P_{\lambda'l}C_{l,th}^{i\times j}\right) \,,
\end{align*}
where $i,j\in\{95~\mathrm{GHz}, 150~\mathrm{GHz}\}$ and $i\times j$ denotes the cross-spectrum between the $T$ map of $i$ band and the $E$ map of $j$ band. We apply this correction on our SPTpol bandpowers as $ C_\beta^{i\times j} \rightarrow C_\beta^{i\times j}-TE\textrm{bias}^{i\times j}$.

\section{Bandpowers} \label{sec:bandpowers}

\begin{figure*}[htb]
\epsscale{1.125}
\plotone{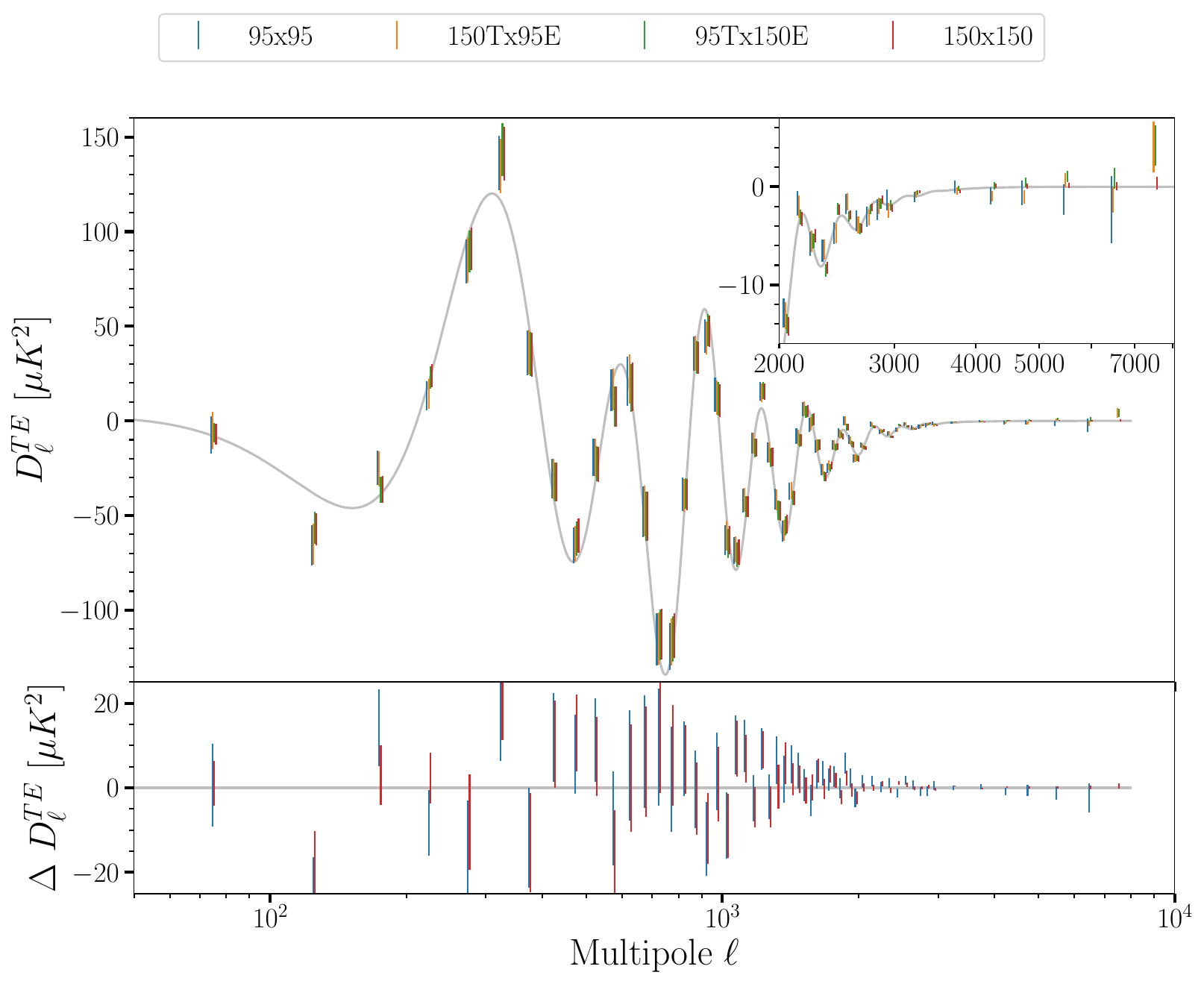}
\caption{SPTpol 500~deg$^2$ $TE$ angular power spectrum, with an inset plot zooming in at high $\ell$. The gray line is the \textit{Planck} best-fit $\Lambda CDM$ model, and the colored error bars are the square root of the diagonal elements of the bandpower covariance matrix without ``further conditioning'' (see Section~\ref{subsec:MVtest}). We plot residuals $\Delta D_\ell$ to the \textit{Planck} best-fit $\Lambda CDM$ model in the subpanel; cross-frequency bandpowers are omitted for better legibility. Small offsets in $\ell$ have been added for plotting purposes.
\label{fig:TE}}
\end{figure*}

\begin{figure*}[htb]
\epsscale{1.125}
\plotone{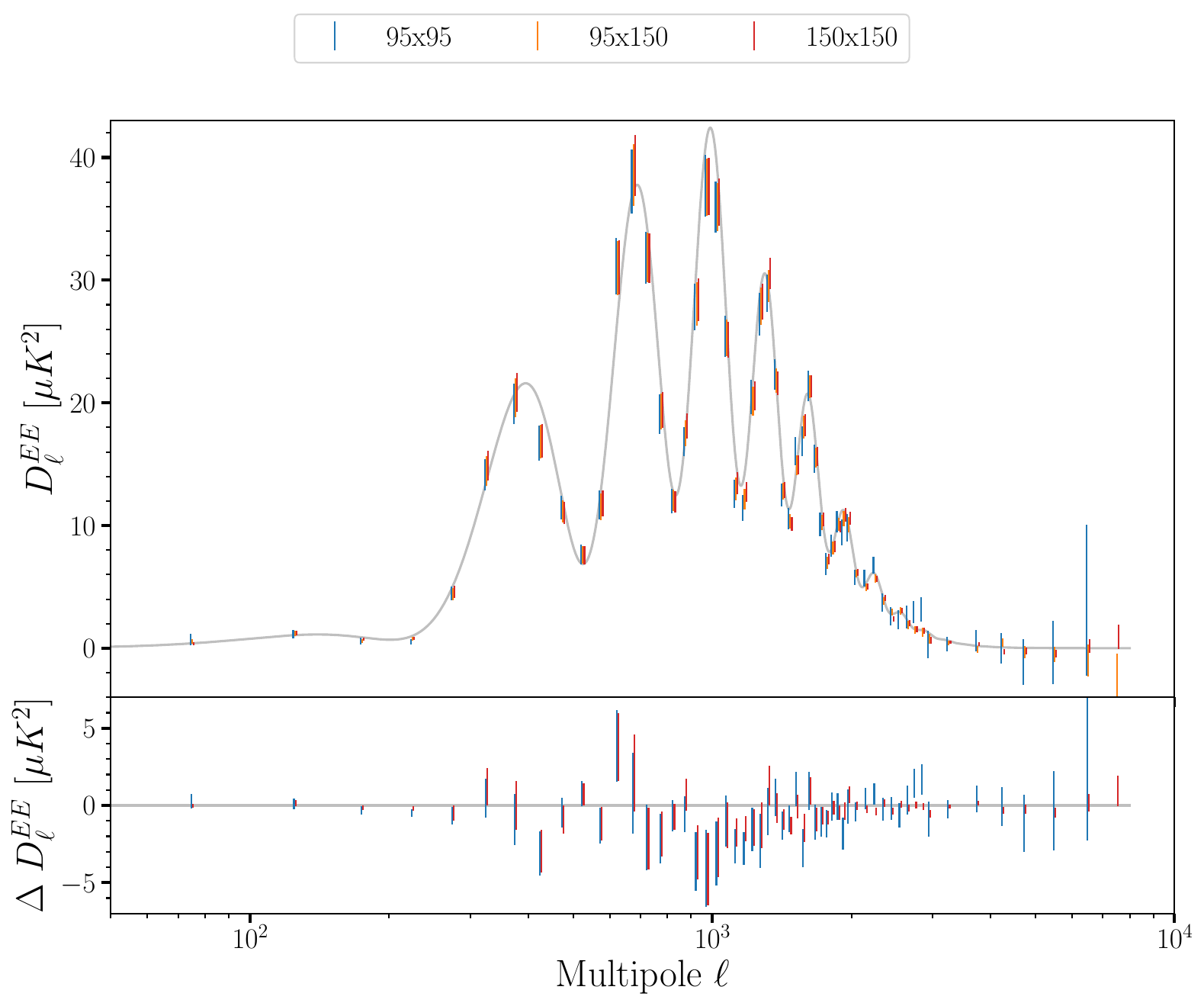}
\caption{SPTpol 500~deg$^2$ $EE$ angular power spectrum. The gray line is the \textit{Planck} best-fit $\Lambda CDM$ model, and the colored error bars are the square root of the diagonal elements of the bandpower covariance matrix without ``further conditioning'' (see Section~\ref{subsec:MVtest}). We plot residuals $\Delta D_\ell$ to the \textit{Planck} best-fit $\Lambda CDM$ model in the subpanel; cross-frequency bandpowers are omitted for better legibility. Small offsets in $\ell$ have been added for plotting purposes.
\label{fig:EE}}
\end{figure*}

\begin{figure*}[htb]
\epsscale{1.2}
\plotone{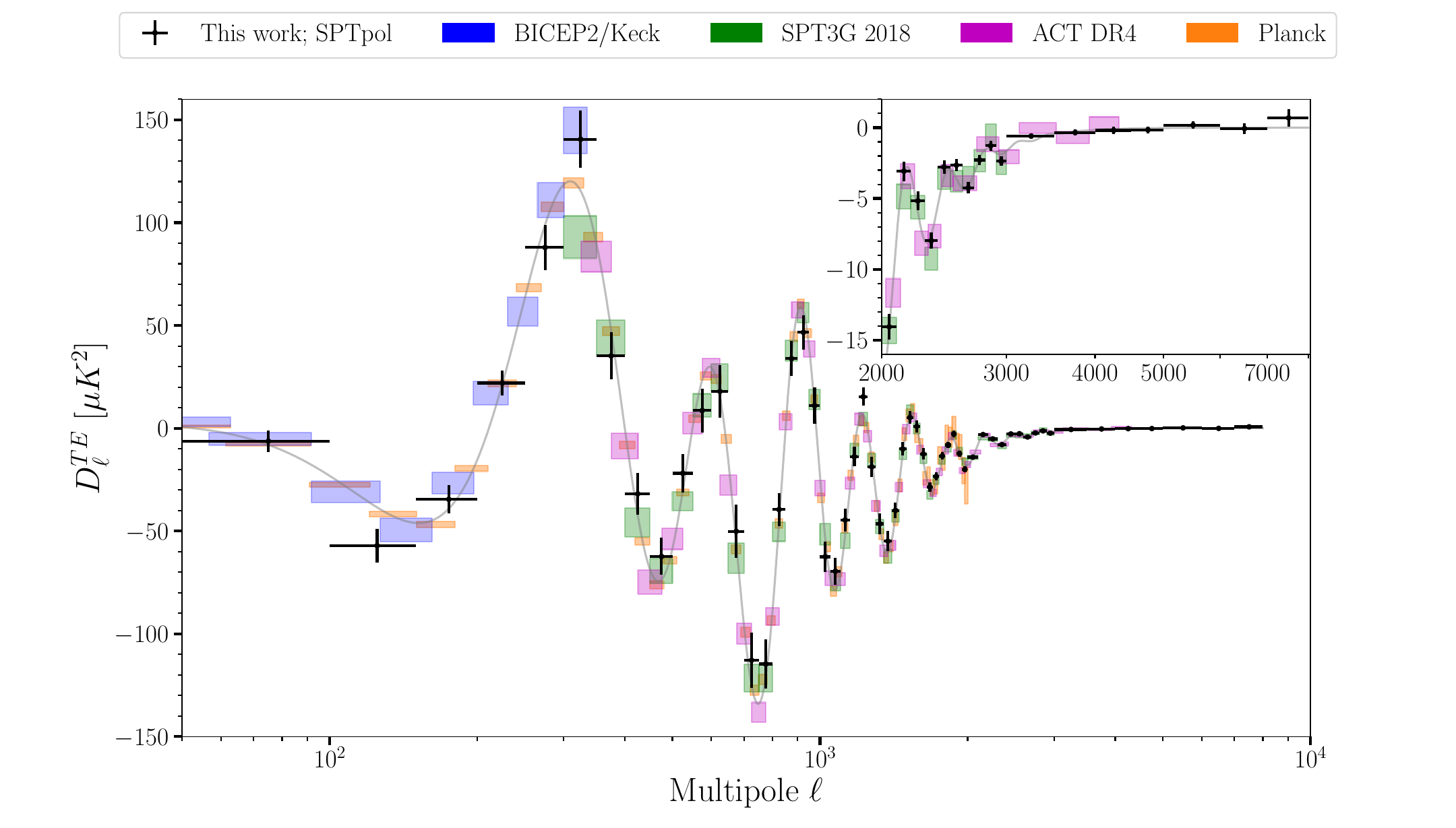}
\caption{SPTpol and other contemporary measurements of the $TE$ angular power spectrum, with an inset plot zooming in at high $\ell$. Black points and error bars are the SPTpol minimum-variance bandpowers in this work. SPT-3G 2018 $TT$/$TE$/$EE$ bandpowers are in green \citep{balkenhol23}, BICEP2/Keck in blue \citep{bicep2keck15}, ACT DR4 in magenta \citep{choi20}, and \textit{Planck} in orange. The gray line is the \textit{Planck} best-fit $\Lambda CDM$ model. 
\label{fig:TE6}}
\end{figure*}

\begin{figure*}[htb]
\epsscale{1.2}
\plotone{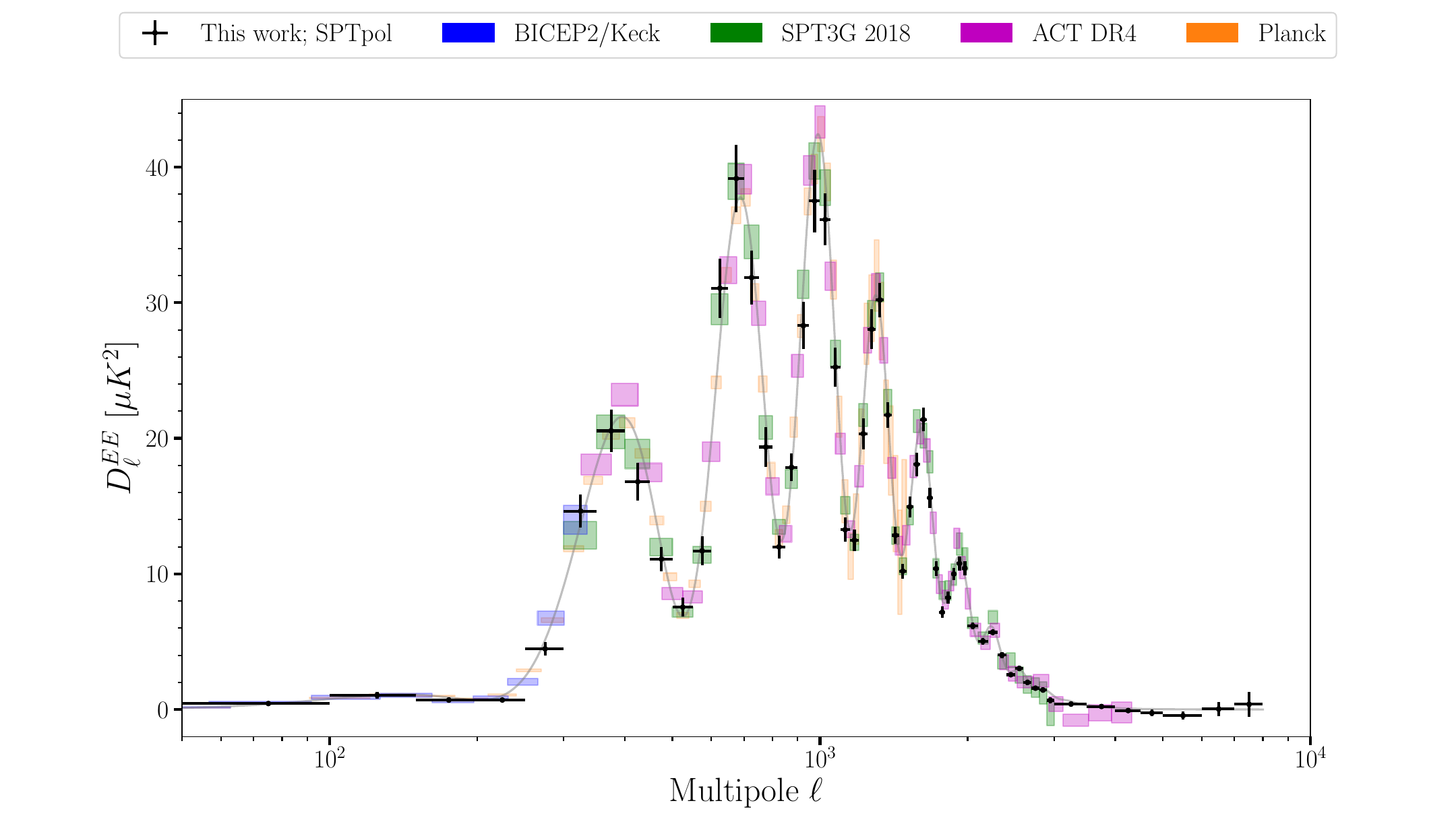}
\caption{SPTpol and other contemporary measurements of the $EE$ angular power spectrum. Black points and error bars are the SPTpol minimum-variance bandpowers in this work. SPT-3G 2018 $TT$/$TE$/$EE$ bandpowers are in green \citep{balkenhol23}, BICEP2/Keck in blue \citep{bicep2keck15}, ACT DR4 in magenta \citep{choi20}, and \textit{Planck} in orange (restricted to $
\ell<1500$). The gray line is the \textit{Planck} best-fit $\Lambda CDM$ model. 
\label{fig:EE7}}
\end{figure*}

Following the procedures in the previous section, we unbias the low-$\ell$ focused power spectra and the high-$\ell$ focused power spectra separately, using the appropriate version of filter transfer function $F$ in each case. The final bandpowers are stitched together using the low-$\ell$ focused power spectra in the range $50<\ell<500$ and the high-$\ell$ focused power spectra in the range $500<\ell<8000$.
These final unbiased bandpowers are shown in Figures~\ref{fig:TE} and~\ref{fig:EE}. The error bars are the square root of the diagonal elements of the bandpower covariance matrix without ``further conditioning'' (see Section~\ref{subsec:MVtest}), and they include contributions from sample variance and noise. We also show our minimum-variance bandpowers alongside other contemporary measurements of the CMB in Figures~\ref{fig:TE6} and~\ref{fig:EE7}.


\subsection{Jackknife Null Tests} \label{subsec:jackknife}

We perform jackknife null tests before the unbiasing step to look for potential systematic contamination. In each test, we sort the 50 map bundles based on that potential systematic, and difference each pair of bundles that are maximally different in that metric. The spectra of these null map bundles are expected to be zero on average, otherwise it indicates our data is contaminated by that systematic.

We perform a total of seven jackknife null tests. The left-right, sun, and moon jackknives are done in the same way as H18. For the azimuth jackknife, in order to maximize sensitivity to azimuth, we no longer use the same 50 map bundles for this test; instead we re-bundle every observation sorted by azimuth into a new set of 50 bundles. We test for potential contamination coming from a building at 153$^\circ$~azimuth by sorting and differencing these new 50 bundles based on distance to 153$^\circ$~azimuth. For the chronological null tests, unlike H18, we perform three separate jackknives: 1st half minus 2nd half within lead-trail observations, 1st half minus 2nd half within full-field observations, and ``1st half of lead-trail plus 1st half of full-field'' minus ``2nd half of lead-trail plus 2nd half of full-field''. This way, for all the jackknives above, their null spectra in signal-only simulations are expected to be small, and we verified that they are indeed negligible, therefore we do not subtract the simulated null spectra from the jackknife null spectra as in H18. Instead, we compute the $\chi^2$ of jackknife null spectra with respect to zero.

In the left-right jackknife, we found excess power in the TE spectra. Looking through each null map bundle, we only saw this excess power in a few date ranges in 2013. We removed observations between 2013 July 27 to August 9, October 25 to November 1, and November 15 to November 27. A total of 228 observations were removed. In the azimuth jackknife, we also found excess power in the lowest $\ell$-bin ($50<\ell<100$) of the 150~GHz spectra.
The source of this power was found to be a coherent azimuth-dependent systematic contamination, concentrated in a small area in 2D Fourier space. 
As a solution, we applied a mask in that area in the 2D Fourier space, for the low-$\ell$ focused dataset only, for both the data maps and simulated maps. 

We present the probabilities to exceed (PTEs) for these jackknife $\chi^2$ tests, calculated after the fixes just discussed, in Table~\ref{tab:jackknife}. 
None of these PTEs fall outside of the threshold typically used in SPT papers \citep[e.g.][]{dutcher21}, namely $0.05 / N_\mathrm{tests}$ or one minus this value.
We conclude that our data does not contain statistically significant systematic biases from the sources tested here.

\begin{table}[h]
\caption{PTE values of each jackknife null test in this work.
\label{tab:jackknife}}
\hspace*{-4em}
\begin{tabular}{|l|ll|ll|}
\hline
                    & \multicolumn{2}{l|}{150~GHz}     & \multicolumn{2}{l|}{95~GHz}      \\ \hline
                    & \multicolumn{1}{l|}{$EE$} & $TE$ & \multicolumn{1}{l|}{$EE$} & $TE$ \\ \hline
Lead-Trail 1st-2nd   & \multicolumn{1}{l|}{0.81} & 0.24 & \multicolumn{1}{l|}{0.70} & 0.39 \\ \hline
Full-Field 1st-2nd   & \multicolumn{1}{l|}{0.29} & 0.31 & \multicolumn{1}{l|}{0.15} & 0.35 \\ \hline
both 1st - both 2nd & \multicolumn{1}{l|}{0.13} & 0.10 & \multicolumn{1}{l|}{0.36} & 0.04 \\ \hline
left-right          & \multicolumn{1}{l|}{0.05} & 0.20 & \multicolumn{1}{l|}{0.28} & 0.18 \\ \hline
sun                 & \multicolumn{1}{l|}{0.17} & 0.27 & \multicolumn{1}{l|}{0.69} & 0.70 \\ \hline
moon                & \multicolumn{1}{l|}{0.59} & 0.81 & \multicolumn{1}{l|}{0.02} & 0.33 \\ \hline
azimuth             & \multicolumn{1}{l|}{0.43} & 0.40 & \multicolumn{1}{l|}{0.23} & 0.38 \\ \hline
\end{tabular}
\end{table}

\subsection{Bandpower Covariance Matrix} \label{subsec:covmat}

In addition to measuring the bandpowers, we also need to know their uncertainties and the correlation of these uncertainties, i.e., the bandpower covariance matrix. It accounts for contributions from both sample variance and noise variance. From our 95 and 150~GHz temperature and $E$-mode maps, there are seven possible sets of $TE$ or $EE$ bandpowers one can calculate ($95T\times95E$, $95T\times150E$, $150T\times95E$, $150T\times150E$, $95E\times95E$, $95E\times150E$, $150E\times150E$); we concatenate these seven sets of bandpowers, and therefore our covariance matrix has a $7\times7$ block structure.

H18 added noise map realizations to noiseless simulated maps and then calculated the bandpower covariance matrix from them, but in this work we did not make noisy simulated maps. We found the (co)variance of the noise realizations to be numerically noisy at high $\ell$ (especially for 95~GHz), so instead, we use the expected variance in the algebraic form for the noise part. To be specific, the diagonal elements of each block in the covariance matrix are expected to be \citep{lueker10}:
\begin{equation} \label{eq:ABCD}
\Xi^{AB,CD}_{bb} \simeq \frac{1}{\nu_b}\left(\langle D_b^{AC}\rangle \langle D_b^{BD}\rangle + \langle D_b^{AD}\rangle \langle D_b^{BC}\rangle\right) \,,
\end{equation}
where $\nu_b$ is the effective number of degrees of freedom, and $D_b^{AB}$ is the cross-spectrum between maps $A$ and $B$.
We assume each map in $\{A,B,C,D\}$ can be decomposed into signal, identical in all maps, and noise, uncorrelated between maps, so that the cross-spectrum $D_b^{AB} = S_b + N_b^{AA} \delta_{AB},$ where $\delta_{AB}$ is the Kronecker delta function. Thus, if $A,B,C,D$ are all different maps for example, all the noise cross-spectra in Eq.~\ref{eq:ABCD} are assumed to be zero, and only the signal part of the covariance remains:
$\Xi_{bb} = 2 S_b^2 / \nu_b$. As another example, if $A,B,C,D$ are all the same map, the noise parts will be completely correlated, so the expected variance becomes $\Xi_{bb} = (2 S_b^2 + 4 S_b N_b + 2 N_b^2) / \nu_b$.

We split the covariance into a sum of two parts and obtain them differently, the signal-only part ($S^2$) and the noisy part ($S\times N$ and $N^2$). The $S^2$ part is not very noisy numerically, so we calculate it directly from the auto-spectra of 226 noiseless simulated maps, as in \citet{crites15}. In addition, we use Eq.~\ref{eq:ABCD} backwards to obtain $\nu_b$ for use in the next part:
\begin{itemize}
  \item Let A = C = 95~GHz T map, B = D = 95~GHz E~map, compute $\nu_b$ as $\nu_b^{TE95}$;
  \item Let A = C = 150~GHz T map, B = D = 150~GHz E~map, compute $\nu_b$ as $\nu_b^{TE150}$;
  \item Let A = C = 95~GHz E map, B = D = 95~GHz E~map, compute $\nu_b$ as $\nu_b^{EE95}$;
  \item Let A = C = 150~GHz E map, B = D = 150~GHz E~map, compute $\nu_b$ as $\nu_b^{EE150}$;
\end{itemize}

Finally, our estimate of $\nu_b^{TE}$ is the average of $\nu_b^{TE95}$ and $\nu_b^{TE150}$, and our estimate of $\nu_b^{EE}$ is the average of $\nu_b^{EE95}$ and $\nu_b^{EE150}$. For the $S\times N$ and $N^2$ part, we again use Eq.~\ref{eq:ABCD} to obtain $\Xi_{bb}^{AB,CD}$ using the average spectra $S_b$ of noiseless simulations, the average spectra $N_b$ of noise realizations, and the aforementioned $\nu_b^{TE}$ and $\nu_b^{EE}$.

The off-diagonal elements of each block in the covariance matrix are expected to reflect mode-coupling from two sources, apodization mask and lensing. These elements are again numerically noisy, so we model them in the following way.

Lensing introduces mode-coupling in a ``checkerboard'' pattern \citep{benoitlevy12}, and it is furthermore amplified and distorted under the flat-sky approximation we use. To estimate this resulting correlation matrix, we take 3,000 pairs of noiseless simulated HEALPix \citep{gorski05} skies (one lensed and one unlensed), and convert them into flat-sky maps under the oblique Lambert azimuthal equal-area projection. We apply the same apodization mask as used for the data, and we compute the $TE$ and $EE$ spectra of the masked flat-sky maps. The correlation matrix estimated using the 3,000 unlensed spectra only contains mode-coupling from the mask, and we model it in a similar way as H18, averaging elements the same distance from the diagonal, and setting all elements greater than $\Delta\ell=50$ from the diagonal to zero. On the other hand, the correlation matrix estimated using the 3,000 lensed spectra contains mode-coupling from both the mask and lensing, therefore we subtract from it the aforementioned correlation matrix of unlensed spectra, and we take the remainder as our model of the lensing ``checkerboard''. We compute these model correlation matrices separately for $TE\times TE$ and $EE\times EE$.

We arrange these $TE\times TE$ or $EE\times EE$ model correlation matrices into the same $7\times7$ block structure; for a $TE\times EE$ off-diagonal block, we take the arithmetic mean of $TE\times TE$ and $EE\times EE$. These correlation matrices are then converted into covariance using the main diagonal of the $7\times7$ bandpower covariance matrix, but for the lensing ``checkerboard'' part, only the $S^2$ part of the main diagonal is used, since noise is expected to have zero contribution. Finally, all the off-diagonal elements from the lensing part and the masking part are added to the bandpower covariance matrix. This matrix is then binned to the same coarse $\ell$ bins as our bandpowers.

The bandpower covariance matrix constructed this way is positive definite. It has $392\times392$ elements, but we note that only 226 independent simulated auto-spectra were used in its construction, leading to numerical instability when inverting (near rank deficiency) as we will see in the internal consistency tests (Section~\ref{subsec:MVtest}). 

Our beam uncertainty comes from 3 sources, and as in H18, we do not incorporate them into the bandpower covariance matrix when performing cosmology fits. Instead, we compute beam uncertainty eigenmodes from each of them, and treat their amplitudes as nuisance parameters during cosmology fits. When performing internal consistency tests however, we indeed add the appropriate contribution from beams into the covariance matrix.

The 1st source of beam uncertainty is the variation between the 5 Venus observations. From each Venus observation, we make an instance of $B_\ell^2$, and we make them in the same way for the 95 and 150~GHz bands. To capture the correlation between bands, we concatenate the three types of spectra $(95\times95, 95\times150, 150\times150)$, and compute its 3$\times$3-block covariance matrix. This is the covariance matrix of $B_\ell^2$,  we extract 4 eigenmodes from this matrix, and then under the first-order approximation, we convert it into a bandpower covariance matrix
\begin{equation*}
D_\ell^{i\times j} \; D_{\ell'}^{k\times l} \; \frac{\textrm{cov}\left((B_\ell^2)^{i\times j}, (B_{\ell'}^2)^{k\times l}\right)}{(B_\ell^2)^{i\times j} (B_{\ell'}^2)^{k\times l}}
\end{equation*}
where $D_\ell$ is CMB bandpowers from a fiducial model. The 2nd source is the uncertainty on the size of the pointing jitter kernels. The nominal pointing jitter kernel is estimated by fitting 2D Gaussians to the two brightest point sources in our field, and then averaging over the two. Using the pointing jitter size of each source individually, we compute their $B_\ell^2$ in the same 3-block structure, and the 3$\times$3-block covariance matrix between the two $B_\ell^2$ contains 1 non-trivial eigenmode. We convert it into a bandpower covariance matrix in the same way as described above.

The 3rd source of beam uncertainty comes from the low-$\ell$ beam shape, described by $A_\textrm{beam}^\textrm{S95}$ and $A_\textrm{beam}^\textrm{S150}$ in Section~\ref{subsec:beam}. By definition the beam uncertainty $\delta B_\ell = \delta A_\textrm{beam}\cdot \Delta B_\ell$ where $\delta A_\textrm{beam}$ is the 1-$\sigma$ uncertainty of the $A_\textrm{beam}$ parameter, and we obtain the 1-$\sigma$ uncertainty on $A_\textrm{beam}^{S95}$ during the fits in Section~\ref{subsec:beam} when we allow $A_\textrm{beam}^{S95}$ and $A_\textrm{beam}^{S150}$ to float simultaneously. In that fit we also find the correlation $\textrm{corr}(A_\textrm{beam}^{S95},A_\textrm{beam}^{S150})$ to be $0.45$. From these we can construct bandpower correlation matrices by generalizing the method in \citet{aylor17}. With $i,j,k,l$ denoting either SPTpol 95~GHz or 150~GHz,
\begin{align*}
\rho_{\ell\ell'}^{i\times j,k\times l} =&\; \left\langle \frac{\delta D_\ell}{D_\ell}^{i\times j}\cdot\frac{\delta D_{\ell'}}{D_{\ell'}}^{k\times l} \right\rangle\\
\textrm{where } \frac{\delta D_\ell}{D_\ell}^{i\times j} =&\; \left(1+\frac{\delta B_\ell}{B_\ell}^i\right)^{-1} \left(1+\frac{\delta B_\ell}{B_\ell}^j\right)^{-1}-1\\
\textrm{to first order, \hspace{0.8em}} =&\; -\frac{\delta B_\ell}{B_\ell}^i -\frac{\delta B_\ell}{B_\ell}^j
\end{align*}
\begin{align*}
\Rightarrow \rho_{\ell\ell'}^{i\times j,k\times l} &= \left\langle \frac{\delta B_\ell}{B_\ell}^i\frac{\delta B_{\ell'}}{B_{\ell'}}^k \right\rangle + \left\langle \frac{\delta B_\ell}{B_\ell}^i\frac{\delta B_{\ell'}}{B_{\ell'}}^l \right\rangle \\
&\enspace+ \left\langle \frac{\delta B_\ell}{B_\ell}^j\frac{\delta B_{\ell'}}{B_{\ell'}}^k \right\rangle + \left\langle \frac{\delta B_\ell}{B_\ell}^j\frac{\delta B_{\ell'}}{B_{\ell'}}^l \right\rangle\\
&= \frac{\delta B_\ell}{B_\ell}^i\frac{\delta B_{\ell'}}{B_{\ell'}}^k\textrm{corr}\left(A_\textrm{beam}^i,A_\textrm{beam}^k\right) \\
&\enspace+ \frac{\delta B_\ell}{B_\ell}^i\frac{\delta B_{\ell'}}{B_{\ell'}}^l\textrm{corr}\left(A_\textrm{beam}^i,A_\textrm{beam}^l\right) \\
&\enspace+ \frac{\delta B_\ell}{B_\ell}^j\frac{\delta B_{\ell'}}{B_{\ell'}}^k\textrm{corr}\left(A_\textrm{beam}^j,A_\textrm{beam}^k\right) \\
&\enspace+ \frac{\delta B_\ell}{B_\ell}^j\frac{\delta B_{\ell'}}{B_{\ell'}}^l\textrm{corr}\left(A_\textrm{beam}^j,A_\textrm{beam}^l\right) \,.
\end{align*}

We stack these correlation matrices into a $3\times3$ block structure in the order of $(95\times95, 95\times150, 150\times150)$, and we extract 2 eigenmodes from this matrix. Finally, we use model bandpowers $D_\ell$ to convert it into a bandpower covariance matrix $D_\ell^{i\times j}D_{\ell'}^{k\times l} \rho_{\ell\ell'}^{i\times j,k\times l}$ \citep{dutcher21}. The total contribution from these 3 sources of beam uncertainty to the bandpower covariance matrix is therefore \begin{multline*}
\biggl( \frac{\mathrm{cov}(B_\ell^{2 ,\, i\times j}, B_{\ell'}^{2 ,\, k\times l})_\mathrm{venus} + \mathrm{cov}(B_\ell^{2 ,\, i\times j}, B_{\ell'}^{2 ,\, k\times l})_\mathrm{jitter}}
{B_\ell^{2 ,\, i\times j} B_{\ell'}^{2 ,\, k\times l}} \\
\enspace+ \rho_{\ell\ell'}^{i\times j,k\times l} \biggr) \cdot D_\ell^{i\times j}D_{\ell'}^{k\times l} \,,
\end{multline*}
and the $4+1+2$ beam uncertainty eigenmodes extracted are included in the fitting process described in Section~\ref{subsec:nuisance}, with their amplitudes treated as nuisance parameters. The priors on the beam eigenmode amplitudes are Gaussian, with 1-$\sigma$ widths equal to their eigenvalues.

\subsection{Internal Consistency} \label{subsec:MVtest}

We expect the multifrequency bandpowers to be consistent with each other, assuming differences in foregrounds are small. With this expectation, we can construct minimum-variance
bandpowers $D^{MV}$ in the same way as \citet{mocanu19} and \citet{dutcher21}, and then perform the same chi-squared test to test this expectation.
\begin{equation*}
D^{MV} = \left(\mathbf{X}^\mathrm{T} \mathbf{C}^{-1} \mathbf{X}\right)^{-1} \mathbf{X}^\mathrm{T} \mathbf{C}^{-1} D \,,
\end{equation*}
where $D$ is our 7-block $TE$ and $EE$ bandpowers, $\mathbf{C}$ is our 7$\times$7-block bandpower covariance matrix, and $\mathbf{X}$ is a $392\times112$ design matrix. In the first 56 columns of $\mathbf{X}$, four elements corresponding to $TE$ bandpowers in that $\ell$-bin are 1, others are 0; in the last 56 columns of $\mathbf{X}$, three elements corresponding to $EE$ bandpowers in that $\ell$-bin are 1, others are 0. Next we compute the $\chi^2$ that quantifies the difference between our multi-frequency bandpowers and minimum-variance bandpowers:
\begin{equation*}
\chi^2 = (D-M)^\mathrm{T} \mathbf{C}^{-1} (D-M) \,,
\end{equation*}
where $M = \mathbf{X}D^{MV}$. We find the resulting $\chi^2$ to be too high for 280 degrees of freedom (392 multifrequency bandpowers $-$ 112 minimum-variance bandpowers).

In other words, we do not pass this test when using the bandpower covariance matrix as constructed. It could be due to the near rank deficiency of the bandpower covariance matrix, which is numerically unstable during inversion.
To mitigate that, we choose a threshold for the eigenvalues under which we set them to a large number, essentially erasing information along those eigenmodes. The criterion for choosing this threshold is the $\chi^2$ distribution of the 226 simulated auto-spectra, computed with this conditioned covariance matrix. When $N$ eigenmodes are erased, the distribution of these 226 $\chi^2$ should be consistent with a $\chi^2$ pdf of $392-1-N$ degrees of freedom, and we quantify this using a Kolmogorov-Smirnov test. After trying various cutoff thresholds, we find that a threshold of ``$2.619\times 10^{-5}$ times the largest eigenvalue'' passes all the criteria above. There are 57 eigenvalues smaller than this threshold, and they are set to very large values instead. After this ``further conditioning'', the $\chi^2$ in the minimum-variance test becomes $\chi^2=257.79$ for 223 degrees of freedom (PTE of 0.0548), and the Kolmogorov-Smirnov test gives a PTE of 0.239. As a check that the eigenmodes erased are indeed spurious,  we show that the parameter constraints are not significantly degraded in Section~\ref{subsec:pipelineTest}.

We use this conditioned $7\times7$ bandpower covariance matrix when fitting cosmology with our full data vector in Section~\ref{sec:constraints}. However, the original covariance matrix (before this conditioning) is also saved for later use, as are the 57 eigenvectors that were erased. We will discuss them in Section~\ref{subsec:splits}.

\section{Likelihood} \label{sec:likelihood}

We write the likelihood in the \verb|Cobaya| framework \citep{torrado21} and use Markov Chain Monte Carlo \citep{christensen01,lewis02b} to sample the posterior distributions of the parameters. For CMB power spectra in the $\Lambda CDM$ model, we use the emulator described in~\citet{bolliet23} trained in the \textsc{CosmoPower} framework \citep{spuriomancini22} using outputs from the \verb|CLASS| Boltzmann code \citep{lesgourgues11,blas11}. For the $\Lambda CDM + A_L$ extension, we use the \verb|CAMB| Boltzmann code \citep{lewis99}. The sum of neutrino masses is 0.06~eV in both cases.

\subsection{$\Lambda$CDM Parameters} \label{subsec:priors}

The six $\Lambda CDM$ parameters in the emulator are: the reionization optical depth $\tau$, the Hubble constant $H_0$, the baryon density $\Omega_b h^2$, the dark matter density $\Omega_\mathrm{cdm}h^2$, the scalar spectral index $n_s$, and $\ln(10^{10}A_s)$ where $A_s$ is the amplitude of primordial scalar perturbations at $k=0.05~\mathrm{Mpc}^{-1}$. The emulator also returns the angular size of the sound horizon at recombination ($100\theta_s$) as a derived parameter, but we note that $\theta_s$ in \verb|CLASS| is defined differently than the $\theta_\mathrm{MC}$ defined in \textsc{CosmoMC}, therefore we never directly compare the $\theta_s$ in this work with $\theta_\mathrm{MC}$ in other works.

SPTpol data alone does not have strong constraining power on the reionization optical depth $\tau$. As in \citet{balkenhol23}, we use a \textit{Planck}-based Gaussian prior of $\tau = 0.0540 \pm 0.0074$. For the other five $\Lambda CDM$ parameters, we use uniform priors.

\subsection{Nuisance Parameters} \label{subsec:nuisance}

We fit for the temperature and polarization calibration factors in Section~\ref{subsec:beam}, and we obtained their best-fit values and 1-$\sigma$ uncertainties. Although we already used these best-fit values to calibrate our data bandpowers, here we still allow the calibration factors to float as nuisance parameters, with a Gaussian prior centered on unity and 1-$\sigma$ uncertainties taken from the above. Because we calibrate our 95~GHz data against our 150~GHz data, the $T_\mathrm{cal}$ for 95 and 150~GHz are correlated, therefore we rearrange terms to define four independent nuisance parameters: $T_\mathrm{cal}^{150},P_\mathrm{cal}^{150},T_\mathrm{cal}^\mathrm{95to150},$ and $E_\mathrm{cal}^{95to150}$. $T_\mathrm{cal}^\mathrm{95to150}$ is defined to be $T_\mathrm{cal}^{95}/T_\mathrm{cal}^{150}$, and $E_\mathrm{cal}^{95to150}$ is defined to be $T_\mathrm{cal}^{95}P_\mathrm{cal}^{95}/(T_\mathrm{cal}^{150}P_\mathrm{cal}^{150})$, and they are both independent quantities obtained from the fits in Section~\ref{subsec:beam}. The Gaussian prior 1-$\sigma$ on $T_\mathrm{cal}^\mathrm{95to150}$ is 0.00115, and the 1-$\sigma$ on $E_\mathrm{cal}^\mathrm{95to150}$ is 0.00746. For $T_\mathrm{cal}^{150}$, we also need to incorporate the calibration uncertainty of \textit{Planck}, and the resulting Gaussian prior 1-$\sigma$ is 0.00271. For the polarization calibration factors, their Gaussian prior 1-$\sigma$ are taken from the findings of H18 (0.01).

We extract a total of 7 beam uncertainty eigenmodes when making the bandpower covariance matrix, as mentioned in Section~\ref{subsec:covmat}. In our cosmology fits, we allow the bandpowers to have variations along those 7 eigenmodes $H_\ell^n$,
\begin{align*}
&C_{\ell,n}^\mathrm{beam} = a_\mathrm{beam}^n H_\ell^n\\
&C_\ell \rightarrow C_\ell \prod_{n=1}^7 \left(1+C_{\ell,n}^\mathrm{beam}\right) \,,
\end{align*}
where $n$ indexes the eigenmodes, $C_\ell$ is the theory power spectrum, and $a_\mathrm{beam}^n$ are the nuisance parameters introduced here as the amplitudes of the eigenmodes. We apply Gaussian priors centered on zero to $a_\mathrm{beam}^n$, and the eigenvalues found when we extracted those eigenmodes are used here as the 1-$\sigma$ widths of the priors.

We introduce a total of 8 nuisance parameters for foreground parameterization, and they are defined in a similar way to H18. For the power spectrum of Galactic dust, we follow the model of \citet{planck12-30},
\begin{align*}
D_{\ell,\mathrm{dust}}^{XY95} &= A_{80}^{XY95}\left(\frac{\ell}{80}\right)^{\alpha_{XY}+2}\\
D_{\ell,\mathrm{dust}}^{XY150} &= A_{80}^{XY150}\left(\frac{\ell}{80}\right)^{\alpha_{XY}+2} \,,
\end{align*}
where $XY\in \{TE,EE\}$, the superscripts 95 or 150 denote either frequency band, $A_{80}$ is the amplitude of the spectrum at $\ell=80$ in units of $\mu K^2$, and $\alpha_{XY}$ is the $\ell$-space spectral index. We assume $\alpha_{XY}$ to be the same between 95 and 150~GHz. For the cross-frequency power spectra $95\times150$, we assume dust is 100\% correlated, i.e.
\begin{align*}
D_{\ell,\mathrm{dust}}^{XY95\times150} &= \sqrt{A_{80}^{XY95}\cdot A_{80}^{XY150}} \left(\frac{\ell}{80}\right)^{\alpha_{XY}+2} \,.
\end{align*}
The priors for these nuisance parameters are also motivated by the findings of~\citet{planck12-30}. For all the $A_{80}$ parameters, we use a uniform prior between 0 and 2 $\mu K^2$. For $\alpha_{XY}$, we use a Gaussian prior centered on $-2.42$ with a $\sigma$ of $0.02$. For polarized extragalactic point sources, we use a single component $D_\ell^{EE}\propto \ell^2$ to model the residual power after masking all sources above 50~mJy in unpolarized flux at 95 or 150~GHz. We introduce the nuisance parameters $D_{3000}^{PSEE,95}$ and $D_{3000}^{PSEE,150}$ as the amplitude of this component at $\ell=3000$ for the 95 and 150~GHz $EE$ spectra, respectively. The amplitude of this component for the $95\times150$ $EE$ spectrum is assumed to be $\sqrt{D_{3000}^{PSEE,95}\cdot D_{3000}^{PSEE,150}}$. We apply a uniform prior between 0 and 2 $\mu K^2$ on these two parameters.

As in H18, we use one parameter to account for ``super-sample lensing'' at every step in the Markov chain \citep{manzotti14}:
\begin{equation*}
\hat{C}_\ell^{XY}(\textbf{p};\kappa) = C_\ell^{XY}(\textbf{p}) - \frac{\partial \ell^2C_\ell^{XY}(\textbf{p})}{\partial \ln\ell}\frac{\kappa}{\ell^2} \,,
\end{equation*}
where $\textbf{p}$ is the parameter vector, and the nuisance parameter $\kappa$ is the mean lensing convergence in the field. We apply a Gaussian prior centered on zero with a $\sigma$ of $0.001$ to $\kappa$. For aberration, we apply the usual zero-parameter correction to the theory spectra \citep{jeong14}:
\begin{equation*}
C_\ell \rightarrow C_\ell - C_\ell\frac{d \ln C_\ell}{d \ln\ell}\beta \langle \cos{\theta} \rangle \,,
\end{equation*}
where $\beta = 0.0012309$ and $\langle \cos{\theta} \rangle = -0.40$.

\subsection{Pipeline Consistency Test} \label{subsec:pipelineTest}

We perform several consistency tests on our likelihood and fitting pipeline. First, we try to recover the $\Lambda CDM$ parameters of the input cosmology chosen for making simulated skies in Section~\ref{subsec:xfer}. We take the average of our 226 noise-free simulated bandpowers, thereby reducing sample variance as much as possible, and fit it with the pipeline described in this section.
Using the best-fit values and the marginalized 1-$\sigma$ widths of each parameter, we find the following shifts
compared to the nominal values of the input cosmology (\textit{Planck} \verb|base_plikHM_TT_lowTEB_lensing|): $\Delta \Omega_b h^2=-0.02\sigma, \Delta \Omega_\mathrm{cdm}h^2=0.05\sigma, \Delta H_0=-0.05\sigma, \Delta \tau=0.02\sigma, \Delta(10^9A_s e^{-2\tau}) =0.05\sigma, \Delta n_s=0.01\sigma$. All six parameters are in excellent agreement, with the shifts being smaller than $1/\sqrt{226}$ of a $\sigma$.


Furthermore, when making simulated bandpowers, we also make 90 realizations of an ``alternate cosmology'' as previously mentioned. Our alternate cosmology is the same one as H18 ($\Omega_b h^2=0.018, \Omega_\mathrm{cdm}h^2=0.14, \theta_\mathrm{MC}=1.079, \tau=0.058, A_s=2.2\times10^{-9}, n_s=0.92$), and with foreground amplitudes doubled. We take the average of these alternate cosmology simulated bandpowers, unbias them with the same kernel used in Section~\ref{sec:powspec}, and fit them to $\Lambda CDM$ theory with the same bandpower covariance matrix obtained in Section~\ref{sec:bandpowers}. This allows us to test whether the unbiasing kernel and/or the covariance matrix have a strong dependence on the chosen input cosmology. As a result, we find the following shifts in the parameter constraints: ($\Delta \Omega_b h^2=0.18\sigma, \Delta \Omega_\mathrm{cdm}h^2=-0.52\sigma, \Delta \theta_s=-0.57\sigma, \Delta \tau=-0.01\sigma, \Delta A_s=-0.26\sigma, \Delta n_s=0.11\sigma$). Although this result shows some half-sigma shifts, we note that the $\theta_\mathrm{MC}$ in this cosmology is drastically different from the current best constraints obtained by \textit{Planck} ($\theta_\mathrm{MC}=\sim1.041$). The 1-$\sigma$ uncertainty on $\theta_s$ in this result is 0.00114, which means $\theta$ is $\sim 30\sigma$ away from \textit{Planck}. Therefore this is a very stringent test, and it shows our pipeline is quite robust against variations in the chosen input cosmology.

Finally, we verify that the ``further conditioning'' of the bandpower covariance matrix done in Section~\ref{subsec:MVtest} did not degrade our cosmological parameter constraints too much. In that step, we erased information along 57 eigenmodes of the covariance matrix, and the assumption was that most of those eigenmodes were spurious in the first place. As a test, we fit our full data bandpowers to $\Lambda CDM$ cosmology again in the same way, but this time using the bandpower covariance matrix before the further conditioning step. We compare, not the best-fit parameters, but the 1-$\sigma$ widths of the parameter constraints before and after further conditioning. The result is that the constraint on $10^9A_s e^{-2\tau}$ is degraded by 6\%, the constraint on $n_s$ by 4\%, and the constraints on other parameters by less than 2\%. We conclude that this assumption is valid and that ``further conditioning'' is justified.

\section{Cosmological Constraints} \label{sec:constraints}

Table~\ref{tab:param0} shows the combined cosmological parameter constraints in the $\Lambda CDM$ model from the seven sets of multifrequency bandpowers in this SPTpol $EE/TE$ dataset. We find $H_0=70.48\pm2.16$ km~s$^{-1}$~Mpc$^{-1}$, $\Omega_m=0.271\pm0.026$, and $\sigma_8=0.758\pm0.022$. The $\chi^2$ of the best-fit $\Lambda CDM$ theory curve to the SPTpol data bandpowers is $\chi^2 = 356.55$ for 324 degrees of freedom (PTE of 0.10).

\begingroup
\begin{table}[h]
\caption{$\Lambda CDM$ parameter constraints and 68\% uncertainties from the full SPTpol 500~deg$^2$ dataset. The top half shows free $\Lambda CDM$ parameters (with a Gaussian prior on $\tau = 0.0540\pm0.0074$), and the bottom half shows derived parameters. $H_0$ is expressed in units of km~s$^{-1}$~Mpc$^{-1}$.
\label{tab:param0}}
\hspace*{-2.2em}
\begin{tabular}{|l|l|ll|}
\hline
\multirow{5}{*}{free} & 100 $\Omega_b h^2$ & 2.271 & ± 0.041 \\ \cline{2-4} 
 & $\Omega_c h^2$ & 0.1115 & ± 0.0051 \\ \cline{2-4} 
 & $n_s$ & 0.993 & ± 0.024 \\ \cline{2-4} 
 & $H_0$ & 70.48 & ± 2.16 \\ \cline{2-4} 
 & $\ln{(10^{10}A_s)}$ & 2.968 & ± 0.032 \\ \hline
\multirow{4}{*}{derived} & 100 $\theta_s$ & 1.0409 & ± 0.0015 \\ \cline{2-4} 
 & $10^9 A_s e^{-2\tau}$ & 1.754 & ± 0.052 \\ \cline{2-4} 
 & $\sigma_8$ & 0.758 & ± 0.022 \\ \cline{2-4} 
 & $\Omega_m$ & 0.271 & ± 0.026 \\ \hline
\end{tabular}
\end{table}
\endgroup

We find the polarized point source power $D_{3000}^{PSEE,95} < 0.43~\mu K^2$ and $D_{3000}^{PSEE,150} < 0.035~\mu K^2$ at 95\% confidence. This upper limit in 150~GHz seems to be three times better than H18, but we caution against taking it at face value; if we did not use the prior to constrain $D_{3000}^{PSEE,150}$ between 0 and 2, the best fit would actually prefer a negative value ($-0.045~\mu K^2$) due to the random noise in our bandpowers. Therefore our upper limit on $D_{3000}^{PSEE,150}$ is artificially tight.
As for the Galactic dust power, the upper limits at 95\% confidence are 1.21, 0.89, 0.21, and 0.24 $\mu K^2$ for $D_{80,\mathrm{dust}}^{TE95}$, $D_{80,\mathrm{dust}}^{TE150}$, $D_{80,\mathrm{dust}}^{EE95}$, and $D_{80,\mathrm{dust}}^{EE150}$, respectively.

\subsection{Data Splits} \label{subsec:splits}

As another check of the internal consistency of the SPTpol $EE/TE$ dataset, we split the data into several subsets and compare the best-fit cosmological parameters between each subset and the full dataset.
We also compute the $\chi^2$ and PTE to the $\Lambda CDM$ model for each data subset individually.
We test seven different subsets of the data: the 95 and 150~GHz auto-frequency spectra, the $95\times 150$ cross-frequency spectra, the $TE$-only and $EE$-only spectra, and the $\ell<1000$ and $\ell>1000$ bandpowers. We fit each subset to $\Lambda CDM$ cosmology with the same method described in Section~\ref{sec:likelihood}, where only the relevant nuisance parameters are floated, the rest fixed to the best-fit values from the full SPTpol dataset. In order to keep the fitting methodology comparable between the subset and the full dataset, for the case of each subset, we fit the full dataset again, with these same nuisance parameters fixed or floated.

For each subset, we would like to slice the bandpower covariance matrix to keep only the relevant blocks, however it would be mathematically wrong to slice the matrix after the ``further conditioning'' described in Section~\ref{subsec:MVtest}. We instead slice the original covariance matrix, but in order to keep the methodology ``apples-to-apples'' between the subset and the full dataset, we need to appropriately account for the 57 erased eigenmodes here. Therefore, as part of the likelihood code, whenever we subtract the theory $\Lambda CDM$ curve from the SPTpol bandpowers, we take this residual vector, compute its projections with respect to the 57 eigenvectors mentioned in Section~\ref{subsec:MVtest}, and subtract these projections from the residual vector. This effectively erases information along these 57 eigenvectors from the residual vector. Finally, this residual vector is sliced to keep only the relevant sections, and is used with the sliced covariance matrix for fitting.

We compare the best-fit cosmological parameters between each subset and the full dataset under these conditions.
Similar to \citet{dutcher21}, we quantify their consistency with a parameter-level $\chi^2$ and PTE, over the five parameters $\ln(10^{10}A_\mathrm{s}), n_s, \Omega_bh^2, \Omega_ch^2,$ and $100\theta_s$ (5 degrees of freedom). $\tau$ is excluded here because the constraints on it are prior-dominated. We define this $\chi^2$ as:
\begin{equation} \label{eq:param-cov}
\chi^2 = \Delta p^T C_p^{-1} \Delta p ,
\end{equation}
where $\Delta p$ is a vector of the difference in best-fit parameters between the subset and the full dataset, and $C_p$ is the covariance of the parameter differences. From \citet{gratton19}, $C_p$ simply equals the difference of parameter-level covariance matrices between the subset and the full dataset. The results are listed in Table~\ref{tab:consistency}, and all seven data splits are within the central $95\%$ confidence interval [$2.5\%, 97.5\%$].

\begin{table}[h]
\caption{Parameter-level $\chi^2$ and the associated PTE between each data split and the full dataset. The five parameters being compared are $\ln(10^{10}A_\mathrm{s}), n_s, \Omega_bh^2, \Omega_ch^2,$ and $100\theta_s$.
\label{tab:consistency}}
\hspace*{2.5em}
\begin{tabular}{|l|l|l|}
\hline
Subset        & $\chi^2$  & PTE    \\ \hline
95~GHz       & 6.16  & 29.1\% \\ \hline
150~GHz       & 2.80  & 73.1\% \\ \hline
$95\times150$       & 1.83  & 87.2\% \\ \hline
$TE$            & 1.63  & 89.8\% \\ \hline
$EE$            & 2.16  & 82.7\% \\ \hline
$\ell < 1000$ & 9.87  & 7.9\%  \\ \hline
$\ell > 1000$ & 12.57 & 2.8\%  \\ \hline
\end{tabular}
\end{table}

In addition to comparing to the full dataset, we quantify the goodness-of-fit of each data split on its own by computing the $\chi^2$ PTE of their bandpowers with respect to their best-fit $\Lambda CDM$ theory curve. As shown in Table~\ref{tab:goodness}, the goodness-of-fit is generally improved over H18, except for the $TE$-only subset. The $\chi^2$ for the $TE$-only subset is $\chi^2 = 238.11$ for 180 degrees of freedom (PTE of 0.24\%).
Although the origin of this low PTE is not completely understood, we note that if we remove the highest $\ell$-bin ($7000<\ell<8000$) of $95\times95$ $TE$ (which has negligible effect on cosmology) the PTE increases to 1.5\%. Out of 7 tests we expect the lowest PTE to be this low as frequently as 10\% of the time.

\begin{table}[h]
\caption{Goodness-of-fit for each data split. We compute the $\chi^2$ between the bandpowers and their best-fit $\Lambda CDM$ theory curve, then compute their PTE values.
\label{tab:goodness}}
\hspace*{3.5em}
\begin{tabular}{|l|r|}
\hline
Subset & \multicolumn{1}{l|}{PTE} \\ \hline
95~GHz & 10.8\% \\ \hline
150~GHz & 26.0\% \\ \hline
$95\times150$ & 17.3\% \\ \hline
$TE$ & 0.24\% \\ \hline
$EE$ & 84.8\% \\ \hline
$\ell < 1000$ & 30.5\% \\ \hline
$\ell > 1000$ & 10.4\% \\ \hline
\end{tabular}
\end{table}


\subsection{Comparison with H18} \label{subsec:H18}

Given that our analysis shares much of the same data as H18, it’s interesting to compare to those results more directly. Our overall measurements and constraints are largely consistent with H18, with our analysis also making several improvements.
Notably, some weak tensions in H18 have lessened, including the poor fit to the $\Lambda CDM$ model and the difference in the $\Lambda CDM$ parameters fit to the $TE$ and $EE$ bandpowers.
As a more apples-to-apples comparison to H18, we redo the $\Lambda CDM$ fit of our 150~GHz-only subset, without the additional step of projecting with respect to the 57 eigenvectors as described in Section~\ref{subsec:splits}. The fit is indeed good, with $\chi^2 = 115.24$ for 104 degrees of freedom (PTE of 0.21). Further testing shows that this improvement in goodness-of-fit comes from improvements in the bandpowers, not the covariance matrix: When we redo this 150~GHz-only fit with our bandpowers and the covariance matrix of H18, the fit is equally good (PTE of 0.21), but when we redo this fit with our covariance matrix and the bandpowers of H18, the $\chi^2 = 165.77$ for 104 degrees of freedom (PTE of 0.0001).

Two mild tensions at a $\sim2\sigma$ level that have persisted from H18 are: the differences in cosmological parameters fit to the high and low-$\ell$ bandpowers, and a preference for $A_L<1$ (Section~\ref{subsec:A_L}).
H18 found that the $\ell<1000$ subset prefers a low $H_0$ and a high $10^9 A_s e^{-2\tau}$, whereas the $\ell>1000$ subset prefers a high $H_0$ and a low $10^9 A_s e^{-2\tau}$.
We lay out the $\Lambda CDM$ parameter constraints in this work in Table~\ref{tab:params} for various subsets. The parameters in Table~\ref{tab:params} are derived using bandpower covariance matrices without the step of projecting the 57 eigenvectors; we find that for the covariance matrix of the 150~GHz-only subsets, extra projection is unnecessary because it has fewer entries and has no nearly-identical rows as in the full set. We also do not see numerical instability issues with the single-frequency matrices. As such, columns (b) and (d) are directly comparable with Table 4 in H18, and the constraints are indeed similar.

\begin{table*}[htbp]
\caption{$\Lambda CDM$ parameter constraints and 68\% uncertainties from a few data splits. The top half shows free $\Lambda CDM$ parameters (with a Gaussian prior on $\tau = 0.0540\pm0.0074$), and the bottom half shows derived parameters. $H_0$ is expressed in units of km~s$^{-1}$~Mpc$^{-1}$.
\label{tab:params}}
\hspace*{-6em}
\begin{tabular}{|l|l|ll|ll|ll|ll|}
\hline
\multicolumn{1}{|c|}{} & \multicolumn{1}{c|}{} & \multicolumn{2}{c|}{(a)} & \multicolumn{2}{c|}{(b)} & \multicolumn{2}{c|}{(c)} & \multicolumn{2}{c|}{(d)} \\ \hline
 &  & \multicolumn{2}{l|}{$\ell<1000$, 95~GHz} & \multicolumn{2}{l|}{$\ell<1000$, 150~GHz} & \multicolumn{2}{l|}{$\ell>1000$, 95~GHz} & \multicolumn{2}{l|}{$\ell>1000$, 150~GHz} \\ \hline
\multirow{5}{*}{free} & 100 $\Omega_b h^2$ & 2.220 & ± 0.132 & 2.267 & ± 0.125 & 2.334 & ± 0.083 & 2.207 & ± 0.050 \\ \cline{2-10} 
 & $\Omega_c h^2$ & 0.1327 & ± 0.0134 & 0.1225 & ± 0.0101 & 0.1065 & ± 0.0124 & 0.1003 & ± 0.0078 \\ \cline{2-10} 
 & $n_s$ & 0.902 & ± 0.059 & 0.941 & ± 0.047 & 1.037 & ± 0.066 & 1.061 & ± 0.046 \\ \cline{2-10} 
 & $H_0$ & 62.61 & ± 5.42 & 66.89 & ± 4.62 & 73.13 & ± 5.80 & 74.84 & ± 3.70 \\ \cline{2-10} 
 & $\ln{(10^{10}A_s)}$ & 3.020 & ± 0.047 & 3.017 & ± 0.041 & 2.903 & ± 0.065 & 2.902 & ± 0.051 \\ \hline
\multirow{4}{*}{derived} & 100 $\theta_s$ & 1.0406 & ± 0.0026 & 1.0422 & ± 0.0023 & 1.0411 & ± 0.0018 & 1.0416 & ± 0.0012 \\ \cline{2-10} 
 & $10^9 A_s e^{-2\tau}$ & 1.839 & ± 0.081 & 1.833 & ± 0.070 & 1.636 & ± 0.104 & 1.635 & ± 0.081 \\ \cline{2-10} 
 & $\sigma_8$ & 0.818 & ± 0.044 & 0.800 & ± 0.041 & 0.725 & ± 0.057 & 0.712 & ± 0.036 \\ \cline{2-10} 
 & $\Omega_m$ & 0.397 & ± 0.100 & 0.326 & ± 0.066 & 0.244 & ± 0.056 & 0.220 & ± 0.035 \\ \hline
\end{tabular}
\end{table*}

\begin{figure*}[htb]
\plotone{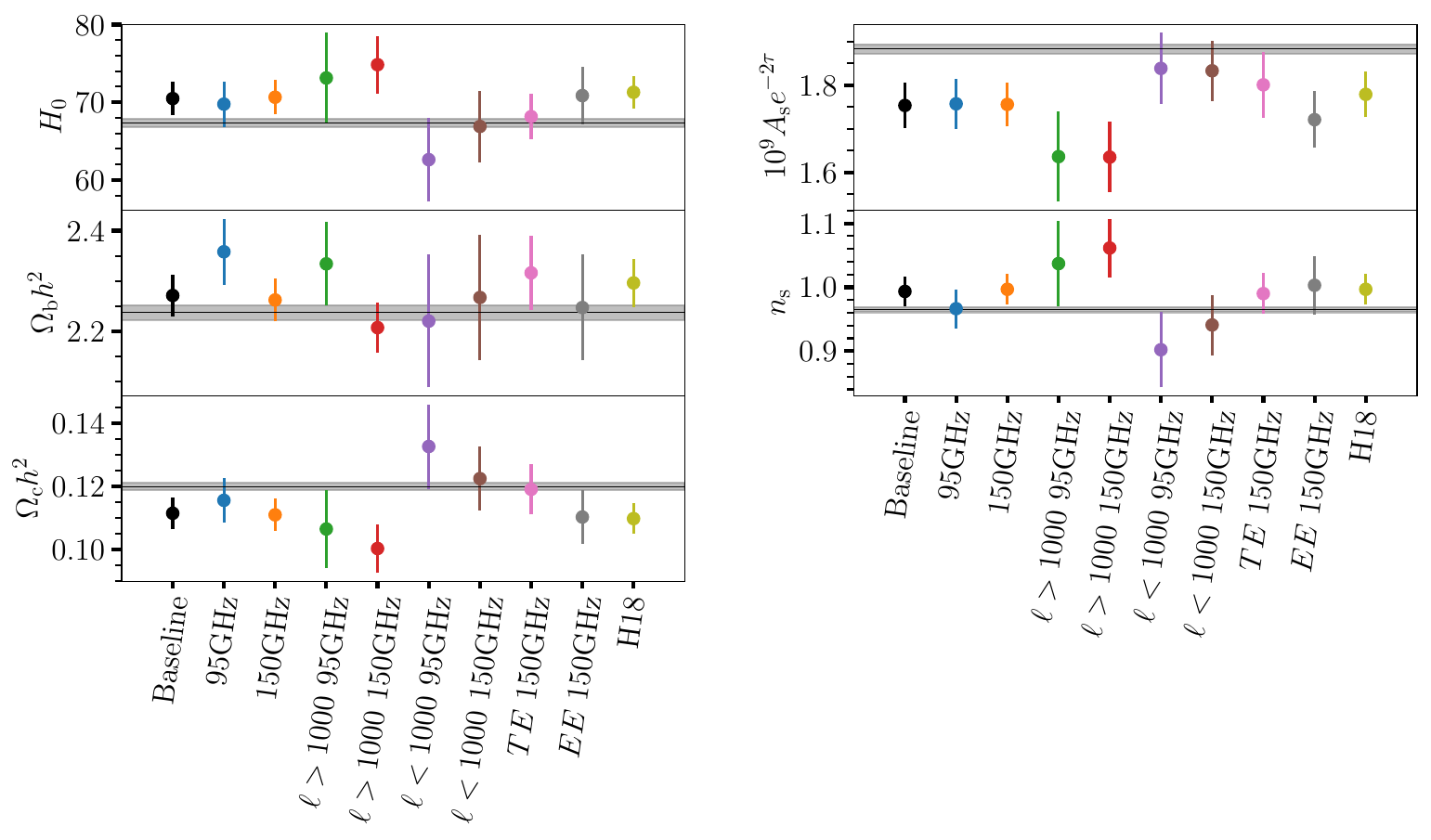}
\caption{$\Lambda CDM$ parameter constraints from the full SPTpol 500 deg$^2$ dataset (baseline), several data splits, and H18. For comparison, the horizontal lines and gray bands are the best-fit values and $1\sigma$ uncertainty ranges of \textit{Planck}. The 150~GHz $TE$ and $EE$ parameter constraints are consistent with each other. The $\ell<1000$ and $\ell>1000$ data splits are borderline consistent at the $\sim2\sigma$ level.
\label{fig:params}}
\end{figure*}

Figure~\ref{fig:params} plots the parameter constraints in Table~\ref{tab:params} and some additional data (sub)sets. We see the same high/low-$\ell$ trend as in H18, in both the 95~GHz-only and the 150~GHz-only subsets.
But as previously noted, the internal consistency between data splits and the full dataset are still within the central $95\%$ confidence interval (the last two rows in Table~\ref{tab:consistency}), therefore this is only a $\lesssim2\sigma$ curiosity.
This trend further weakens in SPT-3G 1500~deg$^2$ analyses \citep{dutcher21,balkenhol23} when data from an additional 1000~deg$^2$ of sky is added, therefore it is unlikely to be a hint for physics beyond $\Lambda CDM$.

\begin{figure*}[htb]
\plotone{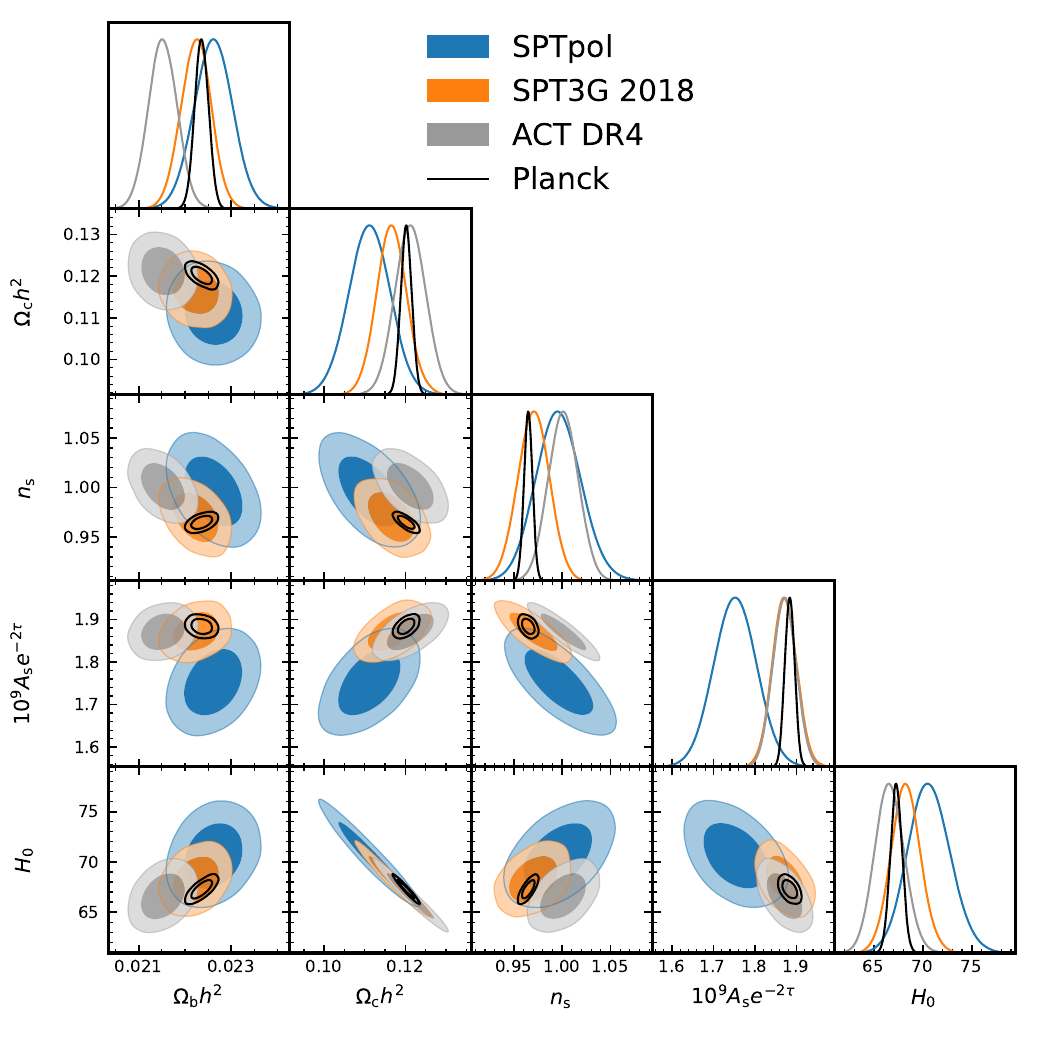}
\caption{Marginalized $\Lambda CDM$ parameter constraints (posteriors) for SPTpol (this work, blue), SPT-3G 2018 $TT$/$TE$/$EE$ (\citealt{balkenhol23}, orange), ACT DR4 (\citealt{aiola20}, gray), and \textit{Planck} (black line contours).
We note that the SPTpol and SPT-3G 2018 constraints are not independent due to shared sample variance (the SPT-3G survey field fully contains the SPTpol field).
\label{fig:tri_ACT}}
\end{figure*}

Alternatively, this trend could be an uncaught systematic bias, but some potential sources of systematics were already ruled out in the jackknife null tests, and some others are being modeled as nuisance parameters.
We can also rule out systematics that affect $TE$ and $EE$ differently, such as T-to-P Leakage, because the $\Lambda CDM$ parameters for $TE$ and $EE$ are consistent with each other and with that of the full dataset (see Table~\ref{tab:consistency}, and data points corresponding to 150~GHz $TE$ and $EE$ in Figure~\ref{fig:params}).
Galactic foregrounds are also unlikely to be contaminating our dataset; as pointed out by H18, the level of $EE$ power from Galactic dust expected in our survey field is a factor of $\sim20$ below our measured $EE$ power in the lowest $\ell$ bin.
One test we perform is to restrict our data to $\ell<3000$, same as SPT-3G analyses, in order to determine whether extragalactic point-source power at high $\ell$ caused these findings to be different between SPT-3G and SPTpol. The result is that the best-fit $\Lambda CDM$ parameters change negligibly, ($\sim0.07\sigma$ in the worst case,) therefore we conclude our model of the point-source power is also adequate.
Lastly, if we multiply the widths of the priors on the beam covariance nuisance parameters by a factor of 10, the $\chi^2$ to the best-fit SPTpol cosmology changes by only $-1.6$. This suggests that our beam uncertainty modes are not driving the cosmological fits.

\subsection{Gravitational Lensing, $A_L$} \label{subsec:A_L}

\begin{figure*}[htb]
\epsscale{0.9}
\plotone{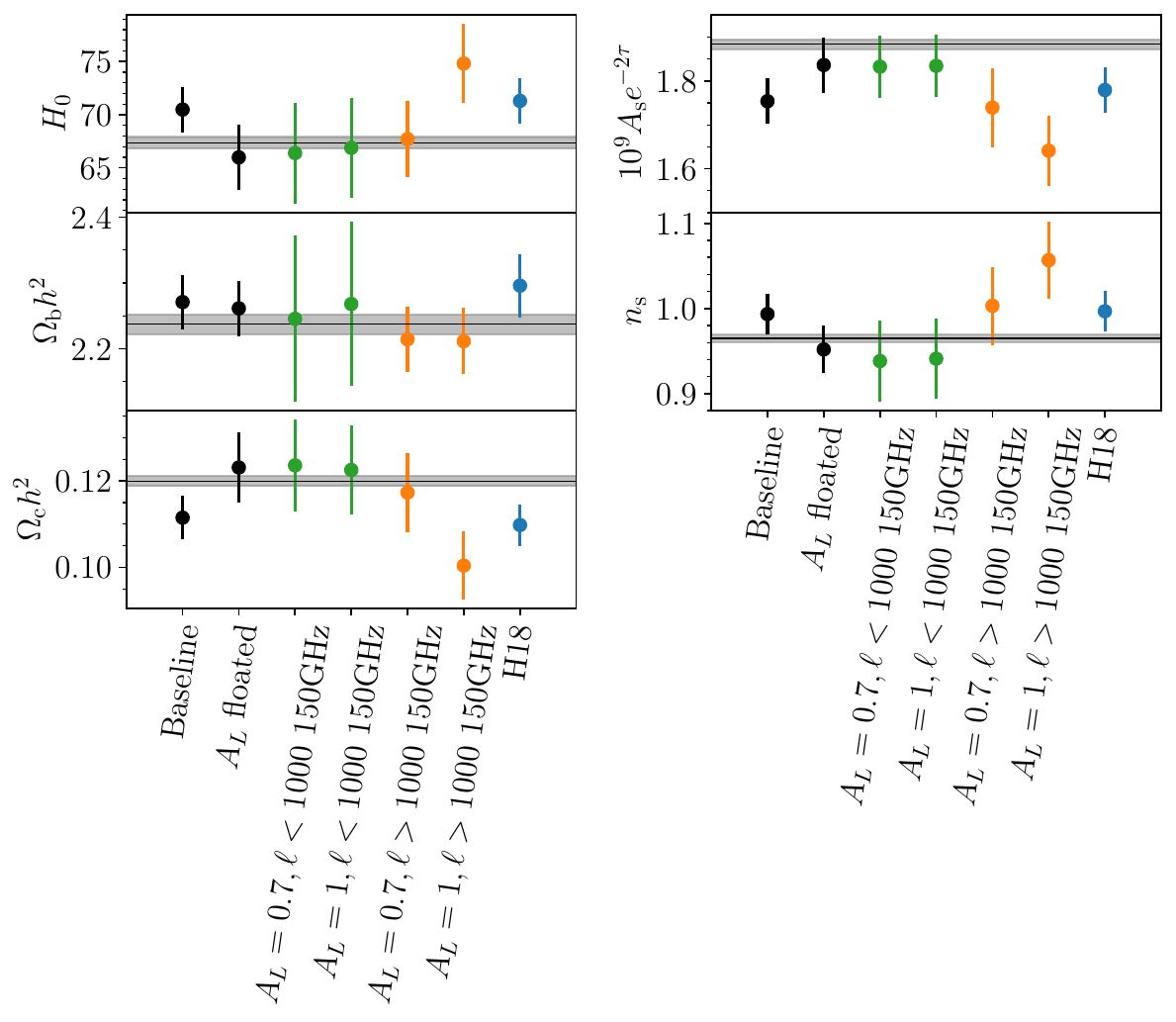}
\caption{Marginalized constraints on $\Lambda CDM$ parameters for the baseline, the full dataset but with $A_L$ floated, the 150~GHz $\ell<1000$ and $\ell>1000$ data splits with $A_L$ fixed to 0.7 or unity, and H18. The horizontal lines and gray bands are the best-fit values and $1\sigma$ uncertainty ranges of \textit{Planck}. When $A_L$ is fixed to 0.7, all these constraints agree between the 150~GHz $\ell<1000$ and $\ell>1000$ data splits, the full dataset, and \textit{Planck}.
\label{fig:Alens}}
\end{figure*}

$A_L$ is the unphysical scaling parameter of the lensing spectrum (see \citet{planck15-13} for details), and H18 found $A_L$ to be $0.81\pm0.14$. We take the same approach and fit the extended model $\Lambda CDM + A_L$ to our full dataset. We find $A_L=0.70\pm0.13$, which is $\sim2\sigma$ lower than unity.

Figure~\ref{fig:tri_ACT} shows the marginalized $\Lambda CDM$ posteriors for this work and other contemporary experiments. These constraints are in good agreement, although the constraint on $10^9A_s e^{-2\tau}$ in this work is $\gtrsim2\sigma$ different from \textit{Planck}. When marginalized over $A_L$, constraints on all $\Lambda CDM$ parameters shift to agree better with \textit{Planck} (black points in Figure~\ref{fig:Alens}). A deeper look at Figure~\ref{fig:params} reveals that the $\ell<1000$ constraints all agree with \textit{Planck}, and that the difference in $10^9A_s e^{-2\tau}$ is primarily driven by the $\ell>1000$ data splits.
CMB lensing not only causes peak smoothing, but transfers power to the damping tail. The $\ell>1000$ data generally favors less structure, a lower matter density, and less CMB lensing, which due to degeneracies in cosmological parameters, also favors larger values for the Hubble constant and scalar tilt (see Table \ref{tab:params}).

By fixing $A_L$ to our best-fit value (0.7) instead of unity, the $\Lambda CDM$ parameter constraints for the 150~GHz $\ell<1000$ data split do not change much, but the constraints for 150~GHz $\ell>1000$ move to agree with the former (see green and orange points in Figure~\ref{fig:Alens}).
When $A_L$ is fixed to 0.7 in this work, the $\Lambda CDM$ constraints all agree between the 150~GHz $\ell<1000$ and $\ell>1000$ data splits, the full dataset, and the nominal \textit{Planck} $\Lambda CDM$ constraints.
We conclude that the differences in cosmological parameters between the high and low-$\ell$ data are driven by features in the high-$\ell$ data that are consistent with less lensing, or lower $A_L$.

One systematic bias that could affect $A_L$ is an uncaught error in the mode-coupling matrix $M_{\lambda\lambda'}$, which would introduce peak smoothing or sharpening. In H18 and in this work, we analytically compute $M_{\lambda\lambda'}$ in the flat-sky regime \citep{hivon02,crites15}. \citet{dutcher21} made $M_{\lambda\lambda'}$ using curved-sky HEALPix simulations, and verified that this simulated $M_{\lambda\lambda'}$ agrees well enough with the analytically-computed $M_{\lambda\lambda'}$ for the 500 deg$^2$ field. Nevertheless, we try swapping in the simulated $M_{\lambda\lambda'}$ in the range $500<\ell<3000$, remaking the 150~GHz $EE$ bandpowers, and rerunning $\Lambda CDM + A_L$ cosmology fits.
The constraint on $A_L$ for the 150~GHz $EE$ subset changed from $0.62\pm0.21$ to $0.59\pm0.20$. We conclude that our $M_{\lambda\lambda'}$ is unlikely to be erroneous enough to drive the preference for $A_L<1$.

In Section~\ref{subsec:splits} we mentioned the $TE$-only data split having a poor goodness-of-fit to $\Lambda CDM$. One might wonder whether the $TE$-only data is driving the preference for $A_L<1$, but that is not the case. Fitting to the $\Lambda CDM+A_L$ extended model, $TE$-only gives $A_L = 0.69 \pm 0.30$, while $EE$-only gives $A_L = 0.61 \pm 0.21$; they both prefer $A_L<1$. In addition, the $TE$-only goodness-of-fit does not improve when the $A_L$ extension is added: It reduces $\chi^2$ by 1.2, and the PTE value remains at 0.24\%. 

In summary, we conclude that the low $A_L$ favored by the SPTpol data is either due to a statistical fluctuation, or a measurement of low lensing power in the SPTpol 500 deg$^2$ patch.
The latter is disfavored by independent measurements of the lensing power that use the lensing-induced correlation between otherwise independent CMB modes \citep{wu19}. The ``lensing reconstruction'' analysis for this 500~deg$^2$ field found that $A_L~\times~A_L^{\phi\phi}=0.995\pm0.090$ \citep{bianchini20a}, where $A_L^{\phi\phi}$ only scales the lensing power regardless of acoustic peak smearing; this constraint on the combination of these two parameters shows that the lensing power in the 500 deg$^2$ field is consistent with $\Lambda CDM$ expectations.

\section{Conclusion} \label{sec:conclusion}

We have presented the full four-year SPTpol 500 deg$^2$ temperature and $E$-mode polarization maps, in both the 95~GHz and the 150~GHz frequency bands. We have also presented the $TE$ and $EE$ angular power spectra in the multipole range $50<\ell<8000$, and they are the most sensitive measurements of the lensed $TE$ and $EE$ power spectra at $\ell >\;\sim1700$, and $\ell >\;\sim2000$, respectively.

This SPTpol dataset is self-consistent. The various cross-frequency bandpowers are consistent with each other, as evidenced by the chi-squared test on the minimum-variance bandpowers.
The $\Lambda CDM$ parameter constraints are consistent across frequency bands, the $TE$ and $EE$ bandpowers, and with the full dataset.
This dataset is a good fit to the $\Lambda CDM$ model, as are most data splits in this work. Using the full SPTpol dataset and a \textit{Planck}-based prior on the optical depth to reionization, we find $H_0=70.48\pm2.16$~km~s$^{-1}$~Mpc$^{-1}$, $\Omega_m=0.271\pm0.026$, and $\sigma_8=0.758\pm0.022$.

We have made several improvements in the analysis that have reduced several weak tensions found in H18, including improvements in the goodness-of-fit to the $\Lambda CDM$ model, and less difference in the $\Lambda CDM$ parameters fit to the $TE$ and $EE$ bandpowers.
The two curiosities that have persisted from H18 are:
The $\Lambda CDM$ parameter constraints from the $\ell<1000$ and $\ell>1000$ data splits are borderline consistent at the $\sim2\sigma$ level, and that the full SPTpol dataset prefers $A_L=0.70\pm0.13$ in the $\Lambda CDM + A_L$ model.

If $A_L$ is fixed to 0.7, the $\ell>1000$ $\Lambda CDM$ parameter constraints shift to agree very well with $\ell<1000$ and \textit{Planck}.
We conclude that the differences in cosmological parameters between the high and low-$\ell$ data are driven by features in the high-$\ell$ data that are consistent with lower $A_L$.
The preference for low $A_L$ in our analysis is more likely due to a statistical fluctuation than an indication of less lensing power in this patch, because the lensing reconstruction analyses from this same data set and field \citep{wu19,bianchini20a} are consistent with $A_L=1$.

We look forward to upcoming SPT-3G and ACT data releases to provide yet higher-sensitivity measurements of the CMB power spectra, as well as measurements from CMB experiments such as the soon-to-be-deployed Simons Observatory Large Aperture Telescope \citep{simonsobservatorycollab19} and CMB-S4 \citep{cmbs4collab19}. In particular, we note the recent powerful cosmological constraints from polarization-only 2019-2020 SPT-3G data \citep{ge24} and the promise of the newly completed SPT-3G Wide Survey (see \citealt{prabhu24} for details). The next few years will bring a wealth of new cosmological information from the CMB.

\section{acknowledgments}

The South Pole Telescope program is supported by the National Science Foundation (NSF) through awards OPP-1852617 and OPP-2332483. Partial support is also provided by the Kavli Institute of Cosmological Physics at the University of Chicago. 

Argonne National Laboratory’s work was supported by the U.S. Department of Energy, Office of High Energy Physics, under contract DE-AC02-06CH11357.
We gratefully acknowledge the computing resources provided on Crossover, a high-performance computing cluster operated by the Laboratory Computing Resource Center at Argonne National Laboratory.

Work at Fermi National Accelerator Laboratory, a DOE-OS, HEP User Facility managed by the Fermi Research Alliance, LLC, was supported under Contract No. DE-AC02-07CH11359.

Melbourne authors acknowledge support from the Australian Research Council’s Discovery Projects scheme (DP210102386).

\section{Data Availability}

Maps, bandpowers, covariance, likelihood, and other supporting data products for this 500-square-degree \textrm{SPTpol} dataset are available on the \href{https://pole.uchicago.edu/public/data/chou25/index.html}{SPT website}.






\bibliography{spt}{}

\begin{thebibliography}{}
\expandafter\ifx\csname natexlab\endcsname\relax\def\natexlab#1{#1}\fi
\providecommand{\url}[1]{\href{#1}{#1}}
\providecommand{\dodoi}[1]{doi:~\href{http://doi.org/#1}{\nolinkurl{#1}}}
\providecommand{\doeprint}[1]{\href{http://ascl.net/#1}{\nolinkurl{http://ascl.net/#1}}}
\providecommand{\doarXiv}[1]{\href{https://arxiv.org/abs/#1}{\nolinkurl{https://arxiv.org/abs/#1}}}

\bibitem[{{Aiola} {et~al.}(2020){Aiola}, {Calabrese}, {Maurin}, {Naess}, {Schmitt}, {Abitbol}, {Addison}, {Ade}, {Alonso}, {Amiri}, {Amodeo}, {Angile}, {Austermann}, {Baildon}, {Battaglia}, {Beall}, {Bean}, {Becker}, {Bond}, {Bruno}, {Calafut}, {Campusano}, {Carrero}, {Chesmore}, {Cho}, {Choi}, {Clark}, {Cothard}, {Crichton}, {Crowley}, {Darwish}, {Datta}, {Denison}, {Devlin}, {Duell}, {Duff}, {Duivenvoorden}, {Dunkley}, {D{\"u}nner}, {Essinger-Hileman}, {Fankhanel}, {Ferraro}, {Fox}, {Fuzia}, {Gallardo}, {Gluscevic}, {Golec}, {Grace}, {Gralla}, {Guan}, {Hall}, {Halpern}, {Han}, {Hargrave}, {Hasselfield}, {Helton}, {Henderson}, {Hensley}, {Hill}, {Hilton}, {Hilton}, {Hincks}, {Hlo{\v{z}}ek}, {Ho}, {Hubmayr}, {Huffenberger}, {Hughes}, {Infante}, {Irwin}, {Jackson}, {Klein}, {Knowles}, {Koopman}, {Kosowsky}, {Lakey}, {Li}, {Li}, {Li}, {Lokken}, {Louis}, {Lungu}, {MacInnis}, {Madhavacheril}, {Maldonado}, {Mallaby-Kay}, {Marsden}, {McMahon}, {Menanteau}, {Moodley}, {Morton}, {Namikawa}, {Nati}, {Newburgh},
  {Nibarger}, {Nicola}, {Niemack}, {Nolta}, {Orlowski-Sherer}, {Page}, {Pappas}, {Partridge}, {Phakathi}, {Pisano}, {Prince}, {Puddu}, {Qu}, {Rivera}, {Robertson}, {Rojas}, {Salatino}, {Schaan}, {Schillaci}, {Sehgal}, {Sherwin}, {Sierra}, {Sievers}, {Sifon}, {Sikhosana}, {Simon}, {Spergel}, {Staggs}, {Stevens}, {Storer}, {Sunder}, {Switzer}, {Thorne}, {Thornton}, {Trac}, {Treu}, {Tucker}, {Vale}, {Van Engelen}, {Van Lanen}, {Vavagiakis}, {Wagoner}, {Wang}, {Ward}, {Wollack}, {Xu}, {Zago}, \& {Zhu}}]{aiola20}
{Aiola}, S., {Calabrese}, E., {Maurin}, L., {et~al.} 2020, \jcap, 2020, 047, \dodoi{10.1088/1475-7516/2020/12/047}

\bibitem[{{Aylor} {et~al.}(2017){Aylor}, {Hou}, {Knox}, {Story}, {Benson}, {Bleem}, {Carlstrom}, {Chang}, {Cho}, {Chown}, {Crawford}, {Crites}, {de Haan}, {Dobbs}, {Everett}, {George}, {Halverson}, {Harrington}, {Holder}, {Holzapfel}, {Hrubes}, {Keisler}, {Lee}, {Leitch}, {Luong-Van}, {Marrone}, {McMahon}, {Meyer}, {Millea}, {Mocanu}, {Mohr}, {Natoli}, {Omori}, {Padin}, {Pryke}, {Reichardt}, {Ruhl}, {Sayre}, {Schaffer}, {Shirokoff}, {Staniszewski}, {Stark}, {Vanderlinde}, {Vieira}, \& {Williamson}}]{aylor17}
{Aylor}, K., {Hou}, Z., {Knox}, L., {et~al.} 2017, \apj, 850, 101, \dodoi{10.3847/1538-4357/aa947b}

\bibitem[{{Balkenhol} {et~al.}(2023){Balkenhol}, {Dutcher}, {Spurio Mancini}, {Doussot}, {Benabed}, {Galli}, {Ade}, {Anderson}, {Ansarinejad}, {Archipley}, {Bender}, {Benson}, {Bianchini}, {Bleem}, {Bouchet}, {Bryant}, {Camphuis}, {Carlstrom}, {Cecil}, {Chang}, {Chaubal}, {Chichura}, {Chou}, {Coerver}, {Crawford}, {Cukierman}, {Daley}, {de Haan}, {Dibert}, {Dobbs}, {Everett}, {Feng}, {Ferguson}, {Foster}, {Gambrel}, {Gardner}, {Goeckner-Wald}, {Gualtieri}, {Guidi}, {Guns}, {Halverson}, {Hivon}, {Holder}, {Holzapfel}, {Hood}, {Huang}, {Knox}, {Korman}, {Kuo}, {Lee}, {Lowitz}, {Lu}, {Millea}, {Montgomery}, {Nakato}, {Natoli}, {Noble}, {Novosad}, {Omori}, {Padin}, {Pan}, {Paschos}, {Prabhu}, {Quan}, {Rahimi}, {Rahlin}, {Reichardt}, {Rouble}, {Ruhl}, {Schiappucci}, {Smecher}, {Sobrin}, {Stark}, {Stephen}, {Suzuki}, {Tandoi}, {Thompson}, {Thorne}, {Tucker}, {Umilta}, {Vieira}, {Wang}, {Whitehorn}, {Wu}, {Yefremenko}, {Young}, {Zebrowski}, \& {SPT-3G Collaboration}}]{balkenhol23}
{Balkenhol}, L., {Dutcher}, D., {Spurio Mancini}, A., {et~al.} 2023, \prd, 108, 023510, \dodoi{10.1103/PhysRevD.108.023510}

\bibitem[{{Benoit-L{\'e}vy} {et~al.}(2012){Benoit-L{\'e}vy}, {Smith}, \& {Hu}}]{benoitlevy12}
{Benoit-L{\'e}vy}, A., {Smith}, K.~M., \& {Hu}, W. 2012, \prd, 86, 123008, \dodoi{10.1103/PhysRevD.86.123008}

\bibitem[{{Bianchini} {et~al.}(2020){Bianchini}, {Wu}, {Ade}, {Anderson}, {Austermann}, {Avva}, {Beall}, {Bender}, {Benson}, {Bleem}, {Carlstrom}, {Chang}, {Chaubal}, {Chiang}, {Citron}, {Moran}, {Crawford}, {Crites}, {de Haan}, {Dobbs}, {Everett}, {Gallicchio}, {George}, {Gilbert}, {Gupta}, {Halverson}, {Harrington}, {Henning}, {Hilton}, {Holder}, {Holzapfel}, {Hrubes}, {Huang}, {Hubmayr}, {Irwin}, {Knox}, {Lee}, {Li}, {Lowitz}, {Manzotti}, {McMahon}, {Meyer}, {Millea}, {Mocanu}, {Montgomery}, {Nadolski}, {Natoli}, {Nibarger}, {Noble}, {Novosad}, {Omori}, {Padin}, {Patil}, {Pryke}, {Reichardt}, {Ruhl}, {Saliwanchik}, {Sayre}, {Schaffer}, {Sievers}, {Simard}, {Smecher}, {Stark}, {Story}, {Tucker}, {Vanderlinde}, {Veach}, {Vieira}, {Wang}, {Whitehorn}, \& {Yefremenko}}]{bianchini20a}
{Bianchini}, F., {Wu}, W.~L.~K., {Ade}, P.~A.~R., {et~al.} 2020, \apj, 888, 119, \dodoi{10.3847/1538-4357/ab6082}

\bibitem[{{BICEP2 and Keck Array Collaborations} {et~al.}(2015){BICEP2 and Keck Array Collaborations}, {Ade}, {Ahmed}, {Aikin}, {Alexander}, {Barkats}, {Benton}, {Bischoff}, {Bock}, {Brevik}, {Buder}, {Bullock}, {Buza}, {Connors}, {Crill}, {Dowell}, {Dvorkin}, {Duband}, {Filippini}, {Fliescher}, {Golwala}, {Halpern}, {Harrison}, {Hasselfield}, {Hildebrandt}, {Hilton}, {Hristov}, {Hui}, {Irwin}, {Karkare}, {Kaufman}, {Keating}, {Kefeli}, {Kernasovskiy}, {Kovac}, {Kuo}, {Leitch}, {Lueker}, {Mason}, {Megerian}, {Netterfield}, {Nguyen}, {O'Brient}, {Ogburn}, {Orlando}, {Pryke}, {Reintsema}, {Richter}, {Schwarz}, {Sheehy}, {Staniszewski}, {Sudiwala}, {Teply}, {Thompson}, {Tolan}, {Turner}, {Vieregg}, {Weber}, {Willmert}, {Wong}, \& {Yoon}}]{bicep2keck15}
{BICEP2 and Keck Array Collaborations}, {Ade}, P.~A.~R., {Ahmed}, Z., {et~al.} 2015, \apj, 811, 126, \dodoi{10.1088/0004-637X/811/2/126}

\bibitem[{{Blas} {et~al.}(2011){Blas}, {Lesgourgues}, \& {Tram}}]{blas11}
{Blas}, D., {Lesgourgues}, J., \& {Tram}, T. 2011, \jcap, 2011, 034, \dodoi{10.1088/1475-7516/2011/07/034}

\bibitem[{{Bolliet} {et~al.}(2023){Bolliet}, {Spurio Mancini}, {Hill}, {Madhavacheril}, {Jense}, {Calabrese}, \& {Dunkley}}]{bolliet23}
{Bolliet}, B., {Spurio Mancini}, A., {Hill}, J.~C., {et~al.} 2023, arXiv e-prints, arXiv:2303.01591, \dodoi{10.48550/arXiv.2303.01591}

\bibitem[{{Carlstrom} {et~al.}(2011){Carlstrom}, {Ade}, {Aird}, {Benson}, {Bleem}, {Busetti}, {Chang}, {Chauvin}, {Cho}, {Crawford}, {Crites}, {Dobbs}, {Halverson}, {Heimsath}, {Holzapfel}, {Hrubes}, {Joy}, {Keisler}, {Lanting}, {Lee}, {Leitch}, {Leong}, {Lu}, {Lueker}, {Luong-van}, {McMahon}, {Mehl}, {Meyer}, {Mohr}, {Montroy}, {Padin}, {Plagge}, {Pryke}, {Ruhl}, {Schaffer}, {Schwan}, {Shirokoff}, {Spieler}, {Staniszewski}, {Stark}, {Tucker}, {Vanderlinde}, {Vieira}, \& {Williamson}}]{carlstrom11}
{Carlstrom}, J.~E., {Ade}, P.~A.~R., {Aird}, K.~A., {et~al.} 2011, \pasp, 123, 568, \dodoi{10.1086/659879}

\bibitem[{{Choi} {et~al.}(2020){Choi}, {Hasselfield}, {Ho}, {Koopman}, {Lungu}, {Abitbol}, {Addison}, {Ade}, {Aiola}, {Alonso}, {Amiri}, {Amodeo}, {Angile}, {Austermann}, {Baildon}, {Battaglia}, {Beall}, {Bean}, {Becker}, {Bond}, {Bruno}, {Calabrese}, {Calafut}, {Campusano}, {Carrero}, {Chesmore}, {Cho}, {Clark}, {Cothard}, {Crichton}, {Crowley}, {Darwish}, {Datta}, {Denison}, {Devlin}, {Duell}, {Duff}, {Duivenvoorden}, {Dunkley}, {D{\"u}nner}, {Essinger-Hileman}, {Fankhanel}, {Ferraro}, {Fox}, {Fuzia}, {Gallardo}, {Gluscevic}, {Golec}, {Grace}, {Gralla}, {Guan}, {Hall}, {Halpern}, {Han}, {Hargrave}, {Henderson}, {Hensley}, {Hill}, {Hilton}, {Hilton}, {Hincks}, {Hlo{\v{z}}ek}, {Hubmayr}, {Huffenberger}, {Hughes}, {Infante}, {Irwin}, {Jackson}, {Klein}, {Knowles}, {Kosowsky}, {Lakey}, {Li}, {Li}, {Li}, {Lokken}, {Louis}, {MacInnis}, {Madhavacheril}, {Maldonado}, {Mallaby-Kay}, {Marsden}, {Maurin}, {McMahon}, {Menanteau}, {Moodley}, {Morton}, {Naess}, {Namikawa}, {Nati}, {Newburgh}, {Nibarger}, {Nicola},
  {Niemack}, {Nolta}, {Orlowski-Sherer}, {Page}, {Pappas}, {Partridge}, {Phakathi}, {Prince}, {Puddu}, {Qu}, {Rivera}, {Robertson}, {Rojas}, {Salatino}, {Schaan}, {Schillaci}, {Schmitt}, {Sehgal}, {Sherwin}, {Sierra}, {Sievers}, {Sifon}, {Sikhosana}, {Simon}, {Spergel}, {Staggs}, {Stevens}, {Storer}, {Sunder}, {Switzer}, {Thorne}, {Thornton}, {Trac}, {Treu}, {Tucker}, {Vale}, {Van Engelen}, {Van Lanen}, {Vavagiakis}, {Wagoner}, {Wang}, {Ward}, {Wollack}, {Xu}, {Zago}, \& {Zhu}}]{choi20}
{Choi}, S.~K., {Hasselfield}, M., {Ho}, S.-P.~P., {et~al.} 2020, \jcap, 2020, 045, \dodoi{10.1088/1475-7516/2020/12/045}

\bibitem[{{Christensen} {et~al.}(2001){Christensen}, {Meyer}, {Knox}, \& {Luey}}]{christensen01}
{Christensen}, N., {Meyer}, R., {Knox}, L., \& {Luey}, B. 2001, Classical and Quantum Gravity, 18, 2677.
\newblock \url{http://adsabs.harvard.edu/cgi-bin/nph-bib_query?bibcode=2001CQGra..18.2677C&db_key=PHY}

\bibitem[{{CMB-S4 Collaboration}(2019)}]{cmbs4collab19}
{CMB-S4 Collaboration}. 2019, arXiv e-prints, arXiv:1907.04473.
\newblock \doarXiv{1907.04473}

\bibitem[{{Crites} {et~al.}(2015){Crites}, {Henning}, {Ade}, {Aird}, {Austermann}, {Beall}, {Bender}, {Benson}, {Bleem}, {Carlstrom}, {Chang}, {Chiang}, {Cho}, {Citron}, {Crawford}, {de Haan}, {Dobbs}, {Everett}, {Gallicchio}, {Gao}, {George}, {Gilbert}, {Halverson}, {Hanson}, {Harrington}, {Hilton}, {Holder}, {Holzapfel}, {Hoover}, {Hou}, {Hrubes}, {Huang}, {Hubmayr}, {Irwin}, {Keisler}, {Knox}, {Lee}, {Leitch}, {Li}, {Liang}, {Luong-Van}, {McMahon}, {Mehl}, {Meyer}, {Mocanu}, {Montroy}, {Natoli}, {Nibarger}, {Novosad}, {Padin}, {Pryke}, {Reichardt}, {Ruhl}, {Saliwanchik}, {Sayre}, {Schaffer}, {Smecher}, {Stark}, {Story}, {Tucker}, {Vanderlinde}, {Vieira}, {Wang}, {Whitehorn}, {Yefremenko}, \& {Zahn}}]{crites15}
{Crites}, A.~T., {Henning}, J.~W., {Ade}, P.~A.~R., {et~al.} 2015, \apj, 805, 36, \dodoi{10.1088/0004-637X/805/1/36}

\bibitem[{{Dutcher} {et~al.}(2021){Dutcher}, {Balkenhol}, {Ade}, {Ahmed}, {Anderes}, {Anderson}, {Archipley}, {Avva}, {Aylor}, {Barry}, {Basu Thakur}, {Benabed}, {Bender}, {Benson}, {Bianchini}, {Bleem}, {Bouchet}, {Bryant}, {Byrum}, {Carlstrom}, {Carter}, {Cecil}, {Chang}, {Chaubal}, {Chen}, {Cho}, {Chou}, {Cliche}, {Crawford}, {Cukierman}, {Daley}, {de Haan}, {Denison}, {Dibert}, {Ding}, {Dobbs}, {Everett}, {Feng}, {Ferguson}, {Foster}, {Fu}, {Galli}, {Gambrel}, {Gardner}, {Goeckner-Wald}, {Gualtieri}, {Guns}, {Gupta}, {Guyser}, {Halverson}, {Harke-Hosemann}, {Harrington}, {Henning}, {Hilton}, {Hivon}, {Holder}, {Holzapfel}, {Hood}, {Howe}, {Huang}, {Irwin}, {Jeong}, {Jonas}, {Jones}, {Khaire}, {Knox}, {Kofman}, {Korman}, {Kubik}, {Kuhlmann}, {Kuo}, {Lee}, {Leitch}, {Lowitz}, {Lu}, {Meyer}, {Michalik}, {Millea}, {Montgomery}, {Nadolski}, {Natoli}, {Nguyen}, {Noble}, {Novosad}, {Omori}, {Padin}, {Pan}, {Paschos}, {Pearson}, {Posada}, {Prabhu}, {Quan}, {Raghunathan}, {Rahlin}, {Reichardt}, {Riebel}, {Riedel},
  {Rouble}, {Ruhl}, {Sayre}, {Schiappucci}, {Shirokoff}, {Smecher}, {Sobrin}, {Stark}, {Stephen}, {Story}, {Suzuki}, {Thompson}, {Thorne}, {Tucker}, {Umilta}, {Vale}, {Vanderlinde}, {Vieira}, {Wang}, {Whitehorn}, {Wu}, {Yefremenko}, {Yoon}, {Young}, \& {SPT-3G Collaboration}}]{dutcher21}
{Dutcher}, D., {Balkenhol}, L., {Ade}, P.~A.~R., {et~al.} 2021, \prd, 104, 022003, \dodoi{10.1103/PhysRevD.104.022003}

\bibitem[{{Galli} {et~al.}(2014){Galli}, {Benabed}, {Bouchet}, {Cardoso}, {Elsner}, {Hivon}, {Mangilli}, {Prunet}, \& {Wandelt}}]{galli14}
{Galli}, S., {Benabed}, K., {Bouchet}, F., {et~al.} 2014, \prd, 90, 063504, \dodoi{10.1103/PhysRevD.90.063504}

\bibitem[{{Ge} {et~al.}(2024){Ge}, {Millea}, {Camphuis}, {Daley}, {Huang}, {Omori}, {Quan}, {Anderes}, {Anderson}, {Ansarinejad}, {Archipley}, {Balkenhol}, {Benabed}, {Bender}, {Benson}, {Bianchini}, {Bleem}, {Bouchet}, {Bryant}, {Carlstrom}, {Chang}, {Chaubal}, {Chen}, {Chichura}, {Chokshi}, {Chou}, {Coerver}, {Crawford}, {de Haan}, {Dibert}, {Dobbs}, {Doohan}, {Doussot}, {Dutcher}, {Everett}, {Feng}, {Ferguson}, {Fichman}, {Foster}, {Galli}, {Gambrel}, {Gardner}, {Goeckner-Wald}, {Gualtieri}, {Guidi}, {Guns}, {Halverson}, {Hivon}, {Holder}, {Holzapfel}, {Hood}, {Howe}, {Hryciuk}, {K{\'e}ruzor{\'e}}, {Khalife}, {Knox}, {Korman}, {Kornoelje}, {Kuo}, {Lee}, {Levy}, {Lowitz}, {Lu}, {Maniyar}, {Martsen}, {Menanteau}, {Montgomery}, {Nakato}, {Natoli}, {Noble}, {Pan}, {Paschos}, {Phadke}, {Pollak}, {Prabhu}, {Rahimi}, {Rahlin}, {Reichardt}, {Riebel}, {Rouble}, {Ruhl}, {Schiappucci}, {Sobrin}, {Stark}, {Stephen}, {Tandoi}, {Thorne}, {Trendafilova}, {Umilta}, {Vieira}, {Vitrier}, {Wan}, {Whitehorn}, {Wu}, {Young},
  \& {Zebrowski}}]{ge24}
{Ge}, F., {Millea}, M., {Camphuis}, E., {et~al.} 2024, arXiv e-prints, arXiv:2411.06000, \dodoi{10.48550/arXiv.2411.06000}

\bibitem[{{G{\'o}rski} {et~al.}(2005){G{\'o}rski}, {Hivon}, {Banday}, {Wandelt}, {Hansen}, {Reinecke}, \& {Bartelmann}}]{gorski05}
{G{\'o}rski}, K.~M., {Hivon}, E., {Banday}, A.~J., {et~al.} 2005, \apj, 622, 759, \dodoi{10.1086/427976}

\bibitem[{{Gratton} \& {Challinor}(2020)}]{gratton19}
{Gratton}, S., \& {Challinor}, A. 2020, \mnras, 499, 3410, \dodoi{10.1093/mnras/staa2996}

\bibitem[{{Henning} {et~al.}(2012){Henning}, {Ade}, {Aird}, {Austermann}, {Beall}, {Becker}, {Benson}, {Bleem}, {Britton}, {Carlstrom}, {Chang}, {Cho}, {Crawford}, {Crites}, {Datesman}, {de Haan}, {Dobbs}, {Everett}, {Ewall-Wice}, {George}, {Halverson}, {Harrington}, {Hilton}, {Holzapfel}, {Hubmayr}, {Irwin}, {Karfunkle}, {Keisler}, {Kennedy}, {Lee}, {Leitch}, {Li}, {Lueker}, {Marrone}, {McMahon}, {Mehl}, {Meyer}, {Montgomery}, {Montroy}, {Nagy}, {Natoli}, {Nibarger}, {Niemack}, {Novosad}, {Padin}, {Pryke}, {Reichardt}, {Ruhl}, {Saliwanchik}, {Sayre}, {Schaffer}, {Shirokoff}, {Story}, {Tucker}, {Vanderlinde}, {Vieira}, {Wang}, {Williamson}, {Yefremenko}, {Yoon}, \& {Young}}]{henning12}
{Henning}, J.~W., {Ade}, P., {Aird}, K.~A., {et~al.} 2012, in \procspie, Vol. 8452, Society of Photo-Optical Instrumentation Engineers (SPIE) Conference Series, \dodoi{10.1117/12.927172}

\bibitem[{{Henning} {et~al.}(2018){Henning}, {Sayre}, {Reichardt}, {Ade}, {Anderson}, {Austermann}, {Beall}, {Bender}, {Benson}, {Bleem}, {Carlstrom}, {Chang}, {Chiang}, {Cho}, {Citron}, {Corbett Moran}, {Crawford}, {Crites}, {de Haan}, {Dobbs}, {Everett}, {Gallicchio}, {George}, {Gilbert}, {Halverson}, {Harrington}, {Hilton}, {Holder}, {Holzapfel}, {Hoover}, {Hou}, {Hrubes}, {Huang}, {Hubmayr}, {Irwin}, {Keisler}, {Knox}, {Lee}, {Leitch}, {Li}, {Lowitz}, {Manzotti}, {McMahon}, {Meyer}, {Mocanu}, {Montgomery}, {Nadolski}, {Natoli}, {Nibarger}, {Novosad}, {Padin}, {Pryke}, {Ruhl}, {Saliwanchik}, {Schaffer}, {Sievers}, {Smecher}, {Stark}, {Story}, {Tucker}, {Vanderlinde}, {Veach}, {Vieira}, {Wang}, {Whitehorn}, {Wu}, \& {Yefremenko}}]{henning18}
{Henning}, J.~W., {Sayre}, J.~T., {Reichardt}, C.~L., {et~al.} 2018, \apj, 852, 97, \dodoi{10.3847/1538-4357/aa9ff4}

\bibitem[{{Heymans} {et~al.}(2021){Heymans}, {Tr{\"o}ster}, {Asgari}, {Blake}, {Hildebrandt}, {Joachimi}, {Kuijken}, {Lin}, {S{\'a}nchez}, {van den Busch}, {Wright}, {Amon}, {Bilicki}, {de Jong}, {Crocce}, {Dvornik}, {Erben}, {Fortuna}, {Getman}, {Giblin}, {Glazebrook}, {Hoekstra}, {Joudaki}, {Kannawadi}, {K{\"o}hlinger}, {Lidman}, {Miller}, {Napolitano}, {Parkinson}, {Schneider}, {Shan}, {Valentijn}, {Verdoes Kleijn}, \& {Wolf}}]{heymans20}
{Heymans}, C., {Tr{\"o}ster}, T., {Asgari}, M., {et~al.} 2021, \aap, 646, A140, \dodoi{10.1051/0004-6361/202039063}

\bibitem[{{Hivon} {et~al.}(2002){Hivon}, {G{\'o}rski}, {Netterfield}, {Crill}, {Prunet}, \& {Hansen}}]{hivon02}
{Hivon}, E., {G{\'o}rski}, K.~M., {Netterfield}, C.~B., {et~al.} 2002, \apj, 567, 2, \dodoi{10.1086/338126}

\bibitem[{{Hu} \& {Dodelson}(2002)}]{hu02b}
{Hu}, W., \& {Dodelson}, S. 2002, \araa, 40, 171

\bibitem[{{Jeong} {et~al.}(2014){Jeong}, {Chluba}, {Dai}, {Kamionkowski}, \& {Wang}}]{jeong14}
{Jeong}, D., {Chluba}, J., {Dai}, L., {Kamionkowski}, M., \& {Wang}, X. 2014, \prd, 89, 023003, \dodoi{10.1103/PhysRevD.89.023003}

\bibitem[{{Lesgourgues}(2011)}]{lesgourgues11}
{Lesgourgues}, J. 2011, arXiv e-prints, arXiv:1104.2932, \dodoi{10.48550/arXiv.1104.2932}

\bibitem[{{Lewis} \& {Bridle}(2002)}]{lewis02b}
{Lewis}, A., \& {Bridle}, S. 2002, \prd, 66, 103511

\bibitem[{Lewis {et~al.}(2000)Lewis, Challinor, \& Lasenby}]{lewis99}
Lewis, A., Challinor, A., \& Lasenby, A. 2000, Astrophys. J., 538, 473

\bibitem[{{Lueker} {et~al.}(2010){Lueker}, {Reichardt}, {Schaffer}, {Zahn}, {Ade}, {Aird}, {Benson}, {Bleem}, {Carlstrom}, {Chang}, {Cho}, {Crawford}, {Crites}, {de Haan}, {Dobbs}, {George}, {Hall}, {Halverson}, {Holder}, {Holzapfel}, {Hrubes}, {Joy}, {Keisler}, {Knox}, {Lee}, {Leitch}, {McMahon}, {Mehl}, {Meyer}, {Mohr}, {Montroy}, {Padin}, {Plagge}, {Pryke}, {Ruhl}, {Shaw}, {Shirokoff}, {Spieler}, {Stalder}, {Staniszewski}, {Stark}, {Vanderlinde}, {Vieira}, \& {Williamson}}]{lueker10}
{Lueker}, M., {Reichardt}, C.~L., {Schaffer}, K.~K., {et~al.} 2010, \apj, 719, 1045, \dodoi{10.1088/0004-637X/719/2/1045}

\bibitem[{{Manzotti} {et~al.}(2014){Manzotti}, {Hu}, \& {Benoit-L{\'e}vy}}]{manzotti14}
{Manzotti}, A., {Hu}, W., \& {Benoit-L{\'e}vy}, A. 2014, \prd, 90, 023003, \dodoi{10.1103/PhysRevD.90.023003}

\bibitem[{{Mocanu} {et~al.}(2019){Mocanu}, {Crawford}, {Aylor}, {Benson}, {Bleem}, {Carlstrom}, {Chang}, {Cho}, {Chown}, {Crites}, {de Haan}, {Dobbs}, {Everett}, {George}, {Halverson}, {Harrington}, {Henning}, {Holder}, {Holzapfel}, {Hou}, {Hrubes}, {Knox}, {Lee}, {Luong-Van}, {Marrone}, {McMahon}, {Meyer}, {Millea}, {Mohr}, {Natoli}, {Omori}, {Padin}, {Pryke}, {Reichardt}, {Ruhl}, {Sayre}, {Schaffer}, {Shirokoff}, {Staniszewski}, {Stark}, {Story}, {Vanderlinde}, {Vieira}, {Williamson}, \& {Wu}}]{mocanu19}
{Mocanu}, L.~M., {Crawford}, T.~M., {Aylor}, K., {et~al.} 2019, \jcap, 2019, 038, \dodoi{10.1088/1475-7516/2019/07/038}

\bibitem[{{Planck Collaboration} {et~al.}(2016{\natexlab{a}}){Planck Collaboration}, {Adam}, {Ade}, {Aghanim}, {Arnaud}, {Aumont}, {Baccigalupi}, {Banday}, {Barreiro}, {Bartlett}, {Bartolo}, {Battaner}, {Benabed}, {Benoit-L{\'e}vy}, {Bernard}, {Bersanelli}, {Bielewicz}, {Bonaldi}, {Bonavera}, {Bond}, {Borrill}, {Bouchet}, {Boulanger}, {Bracco}, {Bucher}, {Burigana}, {Butler}, {Calabrese}, {Cardoso}, {Catalano}, {Challinor}, {Chamballu}, {Chary}, {Chiang}, {Christensen}, {Clements}, {Colombi}, {Colombo}, {Combet}, {Couchot}, {Coulais}, {Crill}, {Curto}, {Cuttaia}, {Danese}, {Davies}, {Davis}, {de Bernardis}, {de Zotti}, {Delabrouille}, {Delouis}, {D{\'e}sert}, {Dickinson}, {Diego}, {Dolag}, {Dole}, {Donzelli}, {Dor{\'e}}, {Douspis}, {Ducout}, {Dunkley}, {Dupac}, {Efstathiou}, {Elsner}, {En{\ss}lin}, {Eriksen}, {Falgarone}, {Finelli}, {Forni}, {Frailis}, {Fraisse}, {Franceschi}, {Frejsel}, {Galeotta}, {Galli}, {Ganga}, {Ghosh}, {Giard}, {Giraud-H{\'e}raud}, {Gjerl{\o}w}, {Gonz{\'a}lez-Nuevo}, {G{\'o}rski},
  {Gratton}, {Gregorio}, {Gruppuso}, {Guillet}, {Hansen}, {Hanson}, {Harrison}, {Helou}, {Henrot-Versill{\'e}}, {Hern{\'a}ndez-Monteagudo}, {Herranz}, {Hivon}, {Hobson}, {Holmes}, {Huffenberger}, {Hurier}, {Jaffe}, {Jaffe}, {Jewell}, {Jones}, {Juvela}, {Keih{\"a}nen}, {Keskitalo}, {Kisner}, {Kneissl}, {Knoche}, {Knox}, {Krachmalnicoff}, {Kunz}, {Kurki-Suonio}, {Lagache}, {Lamarre}, {Lasenby}, {Lattanzi}, {Lawrence}, {Leahy}, {Leonardi}, {Lesgourgues}, {Levrier}, {Liguori}, {Lilje}, {Linden-V{\o}rnle}, {L{\'o}pez-Caniego}, {Lubin}, {Mac{\'\i}as-P{\'e}rez}, {Maffei}, {Maino}, {Mandolesi}, {Mangilli}, {Maris}, {Martin}, {Mart{\'\i}nez-Gonz{\'a}lez}, {Masi}, {Matarrese}, {Mazzotta}, {Meinhold}, {Melchiorri}, {Mendes}, {Mennella}, {Migliaccio}, {Mitra}, {Miville-Desch{\^e}nes}, {Moneti}, {Montier}, {Morgante}, {Mortlock}, {Moss}, {Munshi}, {Murphy}, {Naselsky}, {Nati}, {Natoli}, {Netterfield}, {N{\o}rgaard-Nielsen}, {Noviello}, {Novikov}, {Novikov}, {Pagano}, {Pajot}, {Paladini}, {Paoletti}, {Partridge}, {Pasian},
  {Patanchon}, {Pearson}, {Perdereau}, {Perotto}, {Perrotta}, {Pettorino}, {Piacentini}, {Piat}, {Pierpaoli}, {Pietrobon}, {Plaszczynski}, {Pointecouteau}, {Polenta}, {Ponthieu}, {Popa}, {Pratt}, {Prunet}, {Puget}, {Rachen}, {Reach}, {Rebolo}, {Remazeilles}, {Renault}, {Renzi}, {Ricciardi}, {Ristorcelli}, {Rocha}, {Rosset}, {Rossetti}, {Roudier}, {Rouill{\'e} d'Orfeuil}, {Rubi{\~n}o-Mart{\'\i}n}, {Rusholme}, {Sandri}, {Santos}, {Savelainen}, {Savini}, {Scott}, {Soler}, {Spencer}, {Stolyarov}, {Stompor}, {Sudiwala}, {Sunyaev}, {Sutton}, {Suur-Uski}, {Sygnet}, {Tauber}, {Terenzi}, {Toffolatti}, {Tomasi}, {Tristram}, {Tucci}, {Tuovinen}, {Valenziano}, {Valiviita}, {Van Tent}, {Vibert}, {Vielva}, {Villa}, {Wade}, {Wandelt}, {Watson}, {Wehus}, {White}, {White}, {Yvon}, {Zacchei}, \& {Zonca}}]{planck12-30}
{Planck Collaboration}, {Adam}, R., {Ade}, P.~A.~R., {et~al.} 2016{\natexlab{a}}, \aap, 586, A133, \dodoi{10.1051/0004-6361/201425034}

\bibitem[{{Planck Collaboration} {et~al.}(2016{\natexlab{b}}){Planck Collaboration}, {Ade}, {Aghanim}, {Arnaud}, {Ashdown}, {Aumont}, {Baccigalupi}, {Banday}, {Barreiro}, {Bartlett}, \& et~al.}]{planck15-13}
{Planck Collaboration}, {Ade}, P.~A.~R., {Aghanim}, N., {et~al.} 2016{\natexlab{b}}, \aap, 594, A13, \dodoi{10.1051/0004-6361/201525830}

\bibitem[{{Planck Collaboration} {et~al.}(2020){Planck Collaboration}, {Aghanim}, {Akrami}, {Ashdown}, {Aumont}, {Baccigalupi}, {Ballardini}, {Banday}, {Barreiro}, {Bartolo}, {Basak}, {Battye}, {Benabed}, {Bernard}, {Bersanelli}, {Bielewicz}, {Bock}, {Bond}, {Borrill}, {Bouchet}, {Boulanger}, {Bucher}, {Burigana}, {Butler}, {Calabrese}, {Cardoso}, {Carron}, {Challinor}, {Chiang}, {Chluba}, {Colombo}, {Combet}, {Contreras}, {Crill}, {Cuttaia}, {de Bernardis}, {de Zotti}, {Delabrouille}, {Delouis}, {Di Valentino}, {Diego}, {Dor{\'e}}, {Douspis}, {Ducout}, {Dupac}, {Dusini}, {Efstathiou}, {Elsner}, {En{\ss}lin}, {Eriksen}, {Fantaye}, {Farhang}, {Fergusson}, {Fernandez-Cobos}, {Finelli}, {Forastieri}, {Frailis}, {Fraisse}, {Franceschi}, {Frolov}, {Galeotta}, {Galli}, {Ganga}, {G{\'e}nova-Santos}, {Gerbino}, {Ghosh}, {Gonz{\'a}lez-Nuevo}, {G{\'o}rski}, {Gratton}, {Gruppuso}, {Gudmundsson}, {Hamann}, {Handley}, {Hansen}, {Herranz}, {Hildebrandt}, {Hivon}, {Huang}, {Jaffe}, {Jones}, {Karakci}, {Keih{\"a}nen},
  {Keskitalo}, {Kiiveri}, {Kim}, {Kisner}, {Knox}, {Krachmalnicoff}, {Kunz}, {Kurki-Suonio}, {Lagache}, {Lamarre}, {Lasenby}, {Lattanzi}, {Lawrence}, {Le Jeune}, {Lemos}, {Lesgourgues}, {Levrier}, {Lewis}, {Liguori}, {Lilje}, {Lilley}, {Lindholm}, {L{\'o}pez-Caniego}, {Lubin}, {Ma}, {Mac{\'\i}as-P{\'e}rez}, {Maggio}, {Maino}, {Mandolesi}, {Mangilli}, {Marcos-Caballero}, {Maris}, {Martin}, {Martinelli}, {Mart{\'\i}nez-Gonz{\'a}lez}, {Matarrese}, {Mauri}, {McEwen}, {Meinhold}, {Melchiorri}, {Mennella}, {Migliaccio}, {Millea}, {Mitra}, {Miville-Desch{\^e}nes}, {Molinari}, {Montier}, {Morgante}, {Moss}, {Natoli}, {N{\o}rgaard-Nielsen}, {Pagano}, {Paoletti}, {Partridge}, {Patanchon}, {Peiris}, {Perrotta}, {Pettorino}, {Piacentini}, {Polastri}, {Polenta}, {Puget}, {Rachen}, {Reinecke}, {Remazeilles}, {Renzi}, {Rocha}, {Rosset}, {Roudier}, {Rubi{\~n}o-Mart{\'\i}n}, {Ruiz-Granados}, {Salvati}, {Sandri}, {Savelainen}, {Scott}, {Shellard}, {Sirignano}, {Sirri}, {Spencer}, {Sunyaev}, {Suur-Uski}, {Tauber}, {Tavagnacco},
  {Tenti}, {Toffolatti}, {Tomasi}, {Trombetti}, {Valenziano}, {Valiviita}, {Van Tent}, {Vibert}, {Vielva}, {Villa}, {Vittorio}, {Wand elt}, {Wehus}, {White}, {White}, {Zacchei}, \& {Zonca}}]{planck18-6}
{Planck Collaboration}, {Aghanim}, N., {Akrami}, Y., {et~al.} 2020, \aap, 641, A6, \dodoi{10.1051/0004-6361/201833910}

\bibitem[{{Polenta} {et~al.}(2005){Polenta}, {Marinucci}, {Balbi}, {de Bernardis}, {Hivon}, {Masi}, {Natoli}, \& {Vittorio}}]{polenta05}
{Polenta}, G., {Marinucci}, D., {Balbi}, A., {et~al.} 2005, \jcap, 11, 1, \dodoi{10.1088/1475-7516/2005/11/001}

\bibitem[{{Prabhu} {et~al.}(2024){Prabhu}, {Raghunathan}, {Millea}, {Lynch}, {Ade}, {Anderes}, {Anderson}, {Ansarinejad}, {Archipley}, {Balkenhol}, {Benabed}, {Bender}, {Benson}, {Bianchini}, {Bleem}, {Bouchet}, {Bryant}, {Camphuis}, {Carlstrom}, {Cecil}, {Chang}, {Chaubal}, {Chichura}, {Chokshi}, {Chou}, {Coerver}, {Crawford}, {Cukierman}, {Daley}, {de Haan}, {Dibert}, {Dobbs}, {Doussot}, {Dutcher}, {Everett}, {Feng}, {Ferguson}, {Fichman}, {Foster}, {Galli}, {Gambrel}, {Gardner}, {Ge}, {Goeckner-Wald}, {Gualtieri}, {Guidi}, {Guns}, {Halverson}, {Hivon}, {Holder}, {Holzapfel}, {Hood}, {Hryciuk}, {Huang}, {K{\'e}ruzor{\'e}}, {Knox}, {Korman}, {Kornoelje}, {Kuo}, {Lee}, {Levy}, {Lowitz}, {Lu}, {Maniyar}, {Menanteau}, {Montgomery}, {Nakato}, {Natoli}, {Noble}, {Novosad}, {Omori}, {Padin}, {Pan}, {Paschos}, {Phadke}, {Pollak}, {Quan}, {Rahimi}, {Rahlin}, {Reichardt}, {Rouble}, {Ruhl}, {Schiappucci}, {Smecher}, {Sobrin}, {Stark}, {Stephen}, {Suzuki}, {Tandoi}, {Thompson}, {Thorne}, {Trendafilova}, {Tucker},
  {Umilta}, {Vitrier}, {Vieira}, {Wan}, {Wang}, {Whitehorn}, {Wu}, {Yefremenko}, {Young}, \& {Zebrowski}}]{prabhu24}
{Prabhu}, K., {Raghunathan}, S., {Millea}, M., {et~al.} 2024, \apj, 973, 4, \dodoi{10.3847/1538-4357/ad5ff1}

\bibitem[{{Sayre} {et~al.}(2012){Sayre}, {Ade}, {Aird}, {Austermann}, {Beall}, {Becker}, {Benson}, {Bleem}, {Britton}, {Carlstrom}, {Chang}, {Cho}, {Crawford}, {Crites}, {Datesman}, {de Haan}, {Dobbs}, {Everett}, {Ewall-Wice}, {George}, {Halverson}, {Harrington}, {Henning}, {Hilton}, {Holzapfel}, {Hubmayr}, {Irwin}, {Karfunkle}, {Keisler}, {Kennedy}, {Lee}, {Leitch}, {Li}, {Lueker}, {Marrone}, {McMahon}, {Mehl}, {Meyer}, {Montgomery}, {Montroy}, {Nagy}, {Natoli}, {Nibarger}, {Niemack}, {Novosad}, {Padin}, {Pryke}, {Reichardt}, {Ruhl}, {Saliwanchik}, {Schaffer}, {Shirokoff}, {Story}, {Tucker}, {Vanderlinde}, {Vieira}, {Wang}, {Williamson}, {Yefremenko}, {Yoon}, \& {Young}}]{sayre12}
{Sayre}, J.~T., {Ade}, P., {Aird}, K.~A., {et~al.} 2012, in \procspie, Vol. 8452, Society of Photo-Optical Instrumentation Engineers (SPIE) Conference Series, \dodoi{10.1117/12.927035}

\bibitem[{{Shaw} {et~al.}(2010){Shaw}, {Nagai}, {Bhattacharya}, \& {Lau}}]{shaw10}
{Shaw}, L.~D., {Nagai}, D., {Bhattacharya}, S., \& {Lau}, E.~T. 2010, \apj, 725, 1452, \dodoi{10.1088/0004-637X/725/2/1452}

\bibitem[{{Simons Observatory Collaboration}(2019)}]{simonsobservatorycollab19}
{Simons Observatory Collaboration}. 2019, \jcap, 2019, 056, \dodoi{10.1088/1475-7516/2019/02/056}

\bibitem[{{Spurio Mancini} {et~al.}(2022){Spurio Mancini}, {Piras}, {Alsing}, {Joachimi}, \& {Hobson}}]{spuriomancini22}
{Spurio Mancini}, A., {Piras}, D., {Alsing}, J., {Joachimi}, B., \& {Hobson}, M.~P. 2022, \mnras, 511, 1771, \dodoi{10.1093/mnras/stac064}

\bibitem[{{Torrado} \& {Lewis}(2021)}]{torrado21}
{Torrado}, J., \& {Lewis}, A. 2021, \jcap, 2021, 057, \dodoi{10.1088/1475-7516/2021/05/057}

\bibitem[{{Tristram} {et~al.}(2005){Tristram}, {Mac{\'{\i}}as-P{\'e}rez}, {Renault}, \& {Santos}}]{tristram05}
{Tristram}, M., {Mac{\'{\i}}as-P{\'e}rez}, J.~F., {Renault}, C., \& {Santos}, D. 2005, \mnras, 358, 833, \dodoi{10.1111/j.1365-2966.2005.08760.x}

\bibitem[{{Verde} {et~al.}(2023){Verde}, {Sch{\"o}neberg}, \& {Gil-Mar{\'\i}n}}]{verde23}
{Verde}, L., {Sch{\"o}neberg}, N., \& {Gil-Mar{\'\i}n}, H. 2023, arXiv e-prints, arXiv:2311.13305, \dodoi{10.48550/arXiv.2311.13305}

\bibitem[{{Wu} {et~al.}(2019){Wu}, {Mocanu}, {Ade}, {Anderson}, {Austermann}, {Avva}, {Beall}, {Bender}, {Benson}, {Bianchini}, {Bleem}, {Carlstrom}, {Chang}, {Chiang}, {Citron}, {Corbett Moran}, {Crawford}, {Crites}, {de Haan}, {Dobbs}, {Everett}, {Gallicchio}, {George}, {Gilbert}, {Gupta}, {Halverson}, {Harrington}, {Henning}, {Hilton}, {Holder}, {Holzapfel}, {Hou}, {Hrubes}, {Huang}, {Hubmayr}, {Irwin}, {Knox}, {Lee}, {Li}, {Lowitz}, {Manzotti}, {McMahon}, {Meyer}, {Millea}, {Montgomery}, {Nadolski}, {Natoli}, {Nibarger}, {Noble}, {Novosad}, {Omori}, {Padin}, {Patil}, {Pryke}, {Reichardt}, {Ruhl}, {Saliwanchik}, {Sayre}, {Schaffer}, {Sievers}, {Simard}, {Smecher}, {Stark}, {Story}, {Tucker}, {Vanderlinde}, {Veach}, {Vieira}, {Wang}, {Whitehorn}, \& {Yefremenko}}]{wu19}
{Wu}, W.~L.~K., {Mocanu}, L.~M., {Ade}, P.~A.~R., {et~al.} 2019, \apj, 884, 70, \dodoi{10.3847/1538-4357/ab4186}

\bibitem[{{Zaldarriaga}(2001)}]{zaldarriaga01}
{Zaldarriaga}, M. 2001, \prd, 64, 103001, \dodoi{10.1103/PhysRevD.64.103001}

\end{thebibliography}
\bibliographystyle{aasjournal}



\end{document}